\documentclass[acmlarge]{acmart}
\AtBeginDocument{%
  }

\setcopyright{rightsretained}
\copyrightyear{2025}
\acmYear{2025}
\acmDOI{XXXXXXXX.XXXXXX}

\acmVolume{0}
\acmNumber{0}
\acmArticle{0}
\acmMonth{3}

\begin{document}

\title{Leveraging the Dynamics of Leadership in Group Recommendation Systems}

\author{Peijin Yu}
\affiliation{%
  \institution{Graduate School of Information Science and Electrical Engineering, Kyushu University}
  \city{Fukuoka}
  \country{Japan}
}

\author{Shin'ichi Konomi}
\affiliation{%
  \institution{Faculty of Arts and Science, Kyushu University}
  \city{Fukuoka}
  \country{Japan}}
\email{konomi@acm.org}

\renewcommand{\shortauthors}{Yu and Konomi}

\begin{abstract}
In the field of group recommendation systems (GRS), effectively addressing the diverse preferences of group members poses a significant challenge. Traditional GRS approaches often aggregate individual preferences into a collective group preference to generate recommendations, which may overlook the intricate interactions between group members. We introduce a novel approach to group recommendation, with a specific focus on small groups sharing common interests. In particular, we present a web-based restaurant recommendation system that enhances user satisfaction by modeling mutual interactions among group members. Drawing inspiration from group decision-making literature and leveraging graph theory, we propose a recommendation algorithm that emphasizes the dynamics of relationships and trust within the group. By representing group members as nodes and their interactions as directed edges, the algorithm captures pairwise relationships to foster consensus and improve the alignment of recommendations with group preferences. This interaction-focused framework ultimately seeks to enhance overall group satisfaction with the recommended choices.
\end{abstract}

\maketitle

\section{Introduction}
In contemporary society, recommender systems play a pivotal role in shaping our daily experiences by efficiently providing personalized recommendations to individual users. These systems have revolutionized decision-making processes across various domains, such as e-commerce, entertainment, and online education, by tailoring suggestions to individual preferences. However, their inherent design focuses on catering to the needs of a single user at a given time.

In real-world scenarios, there are numerous instances where decisions are made collectively, such as choosing a restaurant for a group outing, selecting a movie for a family evening, or planning a travel itinerary. These cases require recommendations that not only consider individual preferences but also strive to accommodate the diverse interests of all group members. To address these unique challenges, Group Recommender Systems (GRS) have emerged as a specialized field of study~\cite{masthoff2011group}.

Recent studies have proposed various algorithms tailored to specific scenarios. For instance, in the restaurant recommendation domain, Asani et al.~\cite{asani2021restaurant} proposed a recommender system that used sentiment analysis to extract user food preferences from reviews and provide personalized suggestions. In the field of movie recommendations, Katarya et al.~\cite{katarya2017effective} introduces a novel movie recommendation system that leverages k-means clustering and the cuckoo search optimization algorithm, demonstrating improved accuracy and efficiency on the Movielens dataset compared to existing approaches. For travel planning, Sojahrood et al.~\cite{sojahrood2021poi} proposed a fuzzy-based POI group recommendation method that improves accuracy by modeling user influence and considering spatial factors like distance and time in group decisions.

In recent years, advancements in Group Recommender Systems (GRS) have addressed challenges in dynamic group settings. For instance, a dynamic fuzzy group recommender system has been developed to handle user preference uncertainty and group member interactions, enhancing recommendation accuracy~\cite{son2024dynamic}. Additionally, a deep reinforcement learning-based group recommendation system with a multi-head attention mechanism has been proposed to dynamically aggregate group members' preferences, improving recommendation effectiveness~\cite{10387595}. These studies highlight the importance of balancing individual preferences and group dynamics, offering innovative solutions for diverse real-world applications.

GRS aim to synthesize the preferences of group members and generate recommendations that seek to maximize collective satisfaction. Despite their promise, achieving unanimous satisfaction in group settings remains a significant challenge due to the diversity in individual tastes and priorities. Recognizing this complexity, recent research has increasingly focused on understanding the intricate dynamics of influence within groups. 

For example, in the Influence-Based Group Recommendation (IBGR)~\cite{nozari2024ibgr}, limitations in existing models such as Hermes~\cite{10.1007/s10844-016-0400-0} were identified. Hermes relied on pre-assigned, fixed weights for group members based solely on theoretical closeness relationships, which failed to capture the actual interpersonal dynamics. IBGR addressed this by introducing a novel approach to dynamically adjust each member's weight based on the influence exerted by group leaders and the evolving relationships within the group.

In this paper, we propose a novel approach to group recommendation leveraging the dynamics of leadership in small groups, and examine its effectiveness through the experiment with a group recommendation system called {\it EATOUT}, which is designed to enhance overall group satisfaction based on the proposed approach. This paper focuses on the following three key contributions: 
\begin{enumerate}
    \item Emphasizing the pivotal role of group leaders in shaping group preferences;
    \item Exploring multifaceted mutual influences among group members to better reflect real-world dynamics;
    \item Incorporating leadership-oriented recommendation strategies to foster effective group interactions.
\end{enumerate}
These contributions collectively aim to provide a robust framework for addressing the unique challenges of group recommendation scenarios.

\section{Related Works}
\label{cha:RW}

\subsection{Group Recommender Systems}

Early Group Recommender Systems primarily analyzed individual user ratings to generate recommendations, using simple aggregation methods like averaging. For example, MusicFX~\cite{mccarthy1998musicfx} recommends music in a sports club by averaging members' ratings, though it assumes equal influence among members, overlooking group dynamics.

To address these limitations, later models incorporated interactive processes to enhance fairness and satisfaction. CATS~\cite{mccarthy2006cats}, a collaborative system for recommending skiing destinations, uses a tabletop interface for sharing preferences and features a ‘Critique’ mechanism to refine suggestions dynamically based on user feedback. Similarly, HOOTLE~\cite{alvarez2016hootle+} facilitates group decision-making through negotiation, supporting both synthesized and collaboratively created preference lists.

These advancements mark a shift from static models to dynamic, interaction-driven systems that better capture the complexities of group decision-making, improving both recommendation quality and user satisfaction.

\subsection{Aggregation Strategies}
Two primary methodologies for generating group recommendations focus on item features~\cite{marquez2018negotiation}. The first aggregates individual user profiles into a single group profile using content-based filtering to match item attributes with aggregated preferences. While straightforward, this approach assumes homogenization of preferences without significant loss of nuance. The second constructs a unified group profile through deliberative discussions, actively incorporating explicit interactions to foster consensus.

However, these methods often overlook implicit group dynamics, such as influence and unspoken agreements, which significantly impact collective decisions. For instance, systems emphasizing group leadership measure influence through metrics like suggestion frequency or prominence in discussions~\cite{nozari2020novel}. While acknowledging leadership, this approach risks favoring individual preferences, potentially reducing fairness and satisfaction.

To address these challenges, consensus-reaching models dynamically adjust member ratings based on leadership influence while balancing diverse viewpoints~\cite{dong2020consensus}. By integrating leadership dynamics and mitigating biases, these models enhance fairness, inclusivity, and group satisfaction.

This shift from static aggregation to interaction-aware systems reflects the growing recognition of group decision-making complexities, aiming for recommendations that are both accurate and equitable.

\subsection{Group Decision Making}

Group decision-making focuses on identifying a collective solution among multiple alternatives and understanding the trust relationships within the group. This process emphasizes three key aspects: analyzing user weights or influences, estimating incomplete preference values, and implementing feedback mechanisms.

Regarding user weights, these are sometimes determined based on social networks. For instance, greater weight might be assigned to a father's influence compared to that of a friend. In the context of feedback mechanisms, certain models adjust a user's preferences according to those they trust, leading to an evolution of preferences until the consensus degree surpasses a predefined threshold.

Recent studies have proposed models that incorporate trust relationships and expert reliability to enhance decision-making effectiveness. For example, a consensus model combining trust relationships among experts and expert reliability in social network group decision-making has been introduced, where a concept named matching degree is proposed to measure expert reliability~\cite{10018326}. Additionally, a dynamic feedback mechanism with an attitudinal consensus threshold has been developed to generate recommendation advice for identified inconsistent experts, aiming to increase consensus while minimizing adjustment costs~\cite{9350221}.

These advancements highlight the importance of integrating trust dynamics and feedback mechanisms to achieve more accurate and satisfactory outcomes in group decision-making processes.

\subsection{Influence-Based Group Recommendation}

The Influence-Based Group Recommendation (IBGR) algorithm by Nozari and Koohi~\cite{nozari2020novel} serves as a baseline for our subsequent experiment. This algorithm integrates trust and similarity metrics to calculate influence weights, which are used to adjust individual ratings and generate group-level recommendations. It incorporates trust, similarity, and individual influence, along with a leader impact factor, to balance group dynamics and individual preferences.

Trust between users \(u\) and \(v\) is calculated by combining two components: (1){\it partnership}, which measures the overlap of commonly rated items, and (2){\it distance}, which quantifies the similarity in their ratings. {\it Partnership} is defined as:
\begin{equation}
    \text{Partnership}_{u,v} = \frac{|I_u \cap I_v|}{|I_u|},
\end{equation}
and {\it distance} is calculated as:
\begin{equation}
    \text{Distance}_{u,v} = \frac{1}{1 + \sqrt{\sum_{i \in I_{u,v}} (r_{u,i} - r_{v,i})^2}}.
\end{equation}
The trust score is then derived as their harmonic mean:
\begin{equation}
    \text{Trust}_{u,v} = \frac{2 \cdot \text{Partnership}_{u,v} \cdot \text{Distance}_{u,v}}{\text{Partnership}_{u,v} + \text{Distance}_{u,v}}.
\end{equation}

Similarity between users is computed using  {\it Pearson Correlation Coefficient (PCC)}, which captures the linear relationship between their ratings:
\begin{equation}
    \text{Similarity}_{u,v} = \frac{\sum_{i \in I_{u,v}} (r_{u,i} - \bar{r}_u)(r_{v,i} - \bar{r}_v)}{\sqrt{\sum_{i \in I_{u,v}} (r_{u,i} - \bar{r}_u)^2} \cdot \sqrt{\sum_{i \in I_{u,v}} (r_{v,i} - \bar{r}_v)^2}}.
\end{equation}

The influence weight of \(u\) on \(v\) is calculated as the harmonic mean of trust and similarity scores. Adjusted ratings for a user incorporate these weights and the contributions of other group members. For group leaders, a leader impact factor is applied to enhance their influence. Finally, the group rating for an item is the average of the adjusted ratings from all members.

Items are ranked by their group ratings, and the top-\(k\) items are selected as the final recommendations.

\section{Supporting Social Interactions in Group Recommendation Systems}
Group members often interact with each other for their
collective decision making. 
As such interactions can be perfomed online using synchronous
and asynchronous interaction-support mechanisms, 
our group recommendation approach leverages the interaction data captured
from a suite of tools for supporting social interactions and group decision making.  

We show how group interaction-support mechanisms
can be integrated into group recommender systems
by introcuding the user interface (UI) components of {\it EATOUT}, the group recommendation system we developed for supporting groups to decide on the places to eat out. 
The user interface is designed to facilitate efficient group interactions. Each UI component is tailored to enhance user engagement and decision-making. 
We next describe the core features of the interface. 

\subsection{Homepage for Bookmarking and Initial Ratings}

Users are first greeted with a {\it Welcome Page} upon accessing the system. In our experiment, this page provided detailed text explaining the experiment's {\it background}, {\it objectives}, {\it process}, and  {\it system functionalities}. Participants were instructed to carefully review the information before proceeding.
Users next access the {\it Login Page} by clicking the {\tt Login} button on the {\it Welcome Page}. 
The login process ensured secure access, and credentials can be managed by the administrator.

\begin{figure}[h!]
    \centering
    \includegraphics[width=0.9\textwidth]{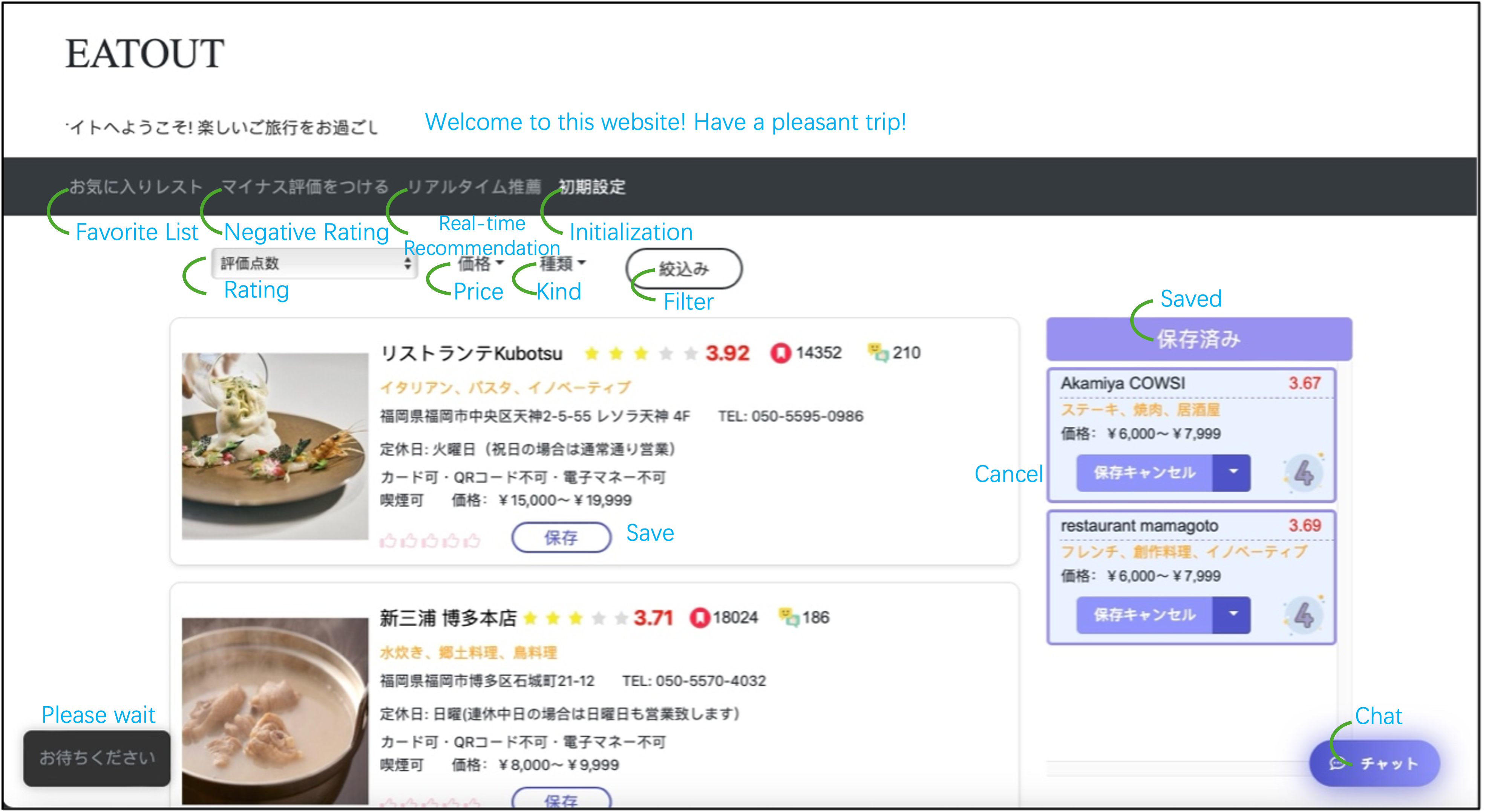}
    \caption{The Homepage Where Users Add Bookmarks and Rate Restaurants.}
    \label{fig:homepage}
\end{figure}

After logging in, users arrive at the {\it Homepage} (see Figure~\ref{fig:homepage}), which allows for the following actions:
\begin{description}
    \item [Bookmark favorite restaurants:] Users could rate and save their preferred restaurants.
    \item [Perform time management:] A countdown timer (bottom-left corner) help users complete a task.
    \item [Access restaurant details:] Clicking on restaurant names allows users to view detailed information.
\end{description}

\subsection{Group Interaction Features}
\begin{figure}[h!]
    \centering
    \includegraphics[width=0.9\textwidth]{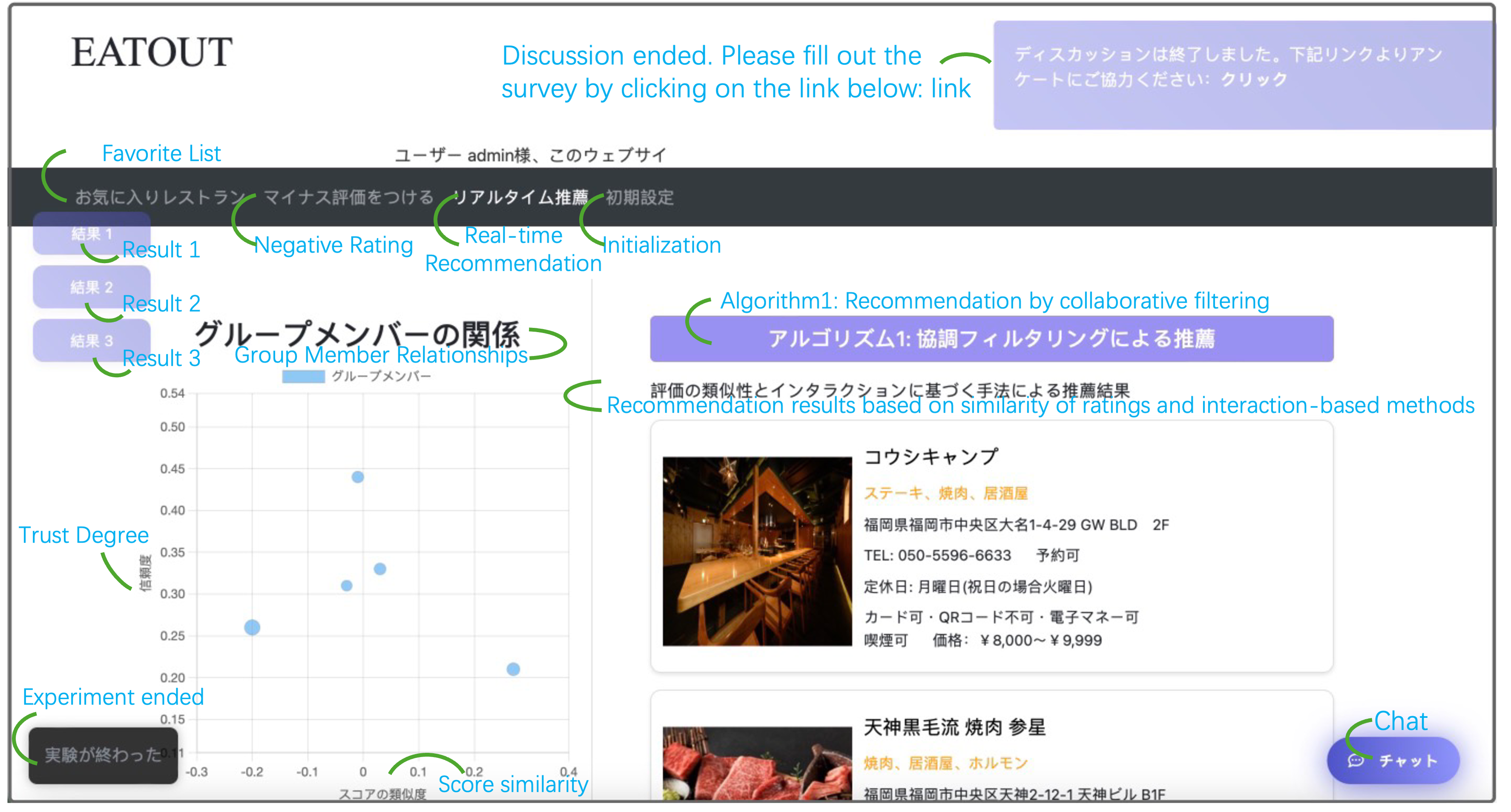}
    \caption{The ``Group Recommendation Results" Page.}
    \label{fig:recommendation_results}
\end{figure}

To facilitate group discussions, EATOUT's user interface includes several interactive features:

\begin{description}
\item[a) Real-Time Chat Panel:] 
Users can use the {\it Real-Time Chat Panel}  by clicking the chat icon in the bottom-right corner of the screen (see Figure~\ref{fig:chat}).
The {\it Real-Time Chat Panel} enables real-time messaging and discussion of preferences. 
When a member share a restaurant they like, it is displayed as a clickable link allowing others to view detailed restaurant information upon clicking. 
It also displays the {\it sender's nickname}, thereby encouraging users to introduce themselves.

\begin{figure}[h!]
    \centering
    \includegraphics[width=0.3\textwidth]{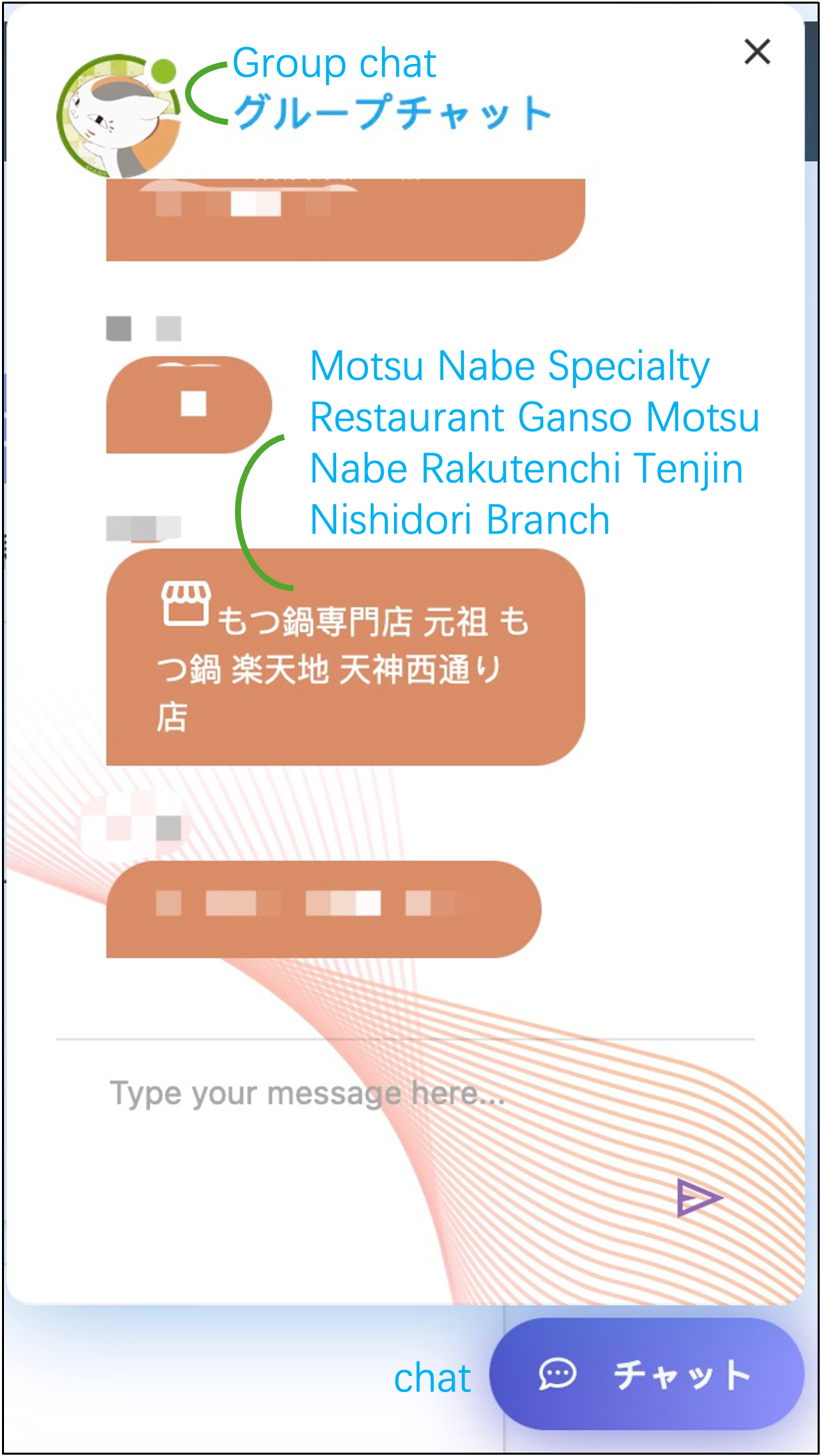}
    \caption{The Real-Time Chat Panel.}
    \label{fig:chat}
\end{figure}

\item[b) Shared Favorite List:]
The {\it Favorite List Page} displays both the restaurants bookmarked by the users themselves and the ones bookmarked by other group members (see Figure~\ref{fig:shared_list}).
Users can select restaurants from other members' lists and add them to their own lists. 

\begin{figure}[h!]
    \centering
    \includegraphics[width=0.9\textwidth]{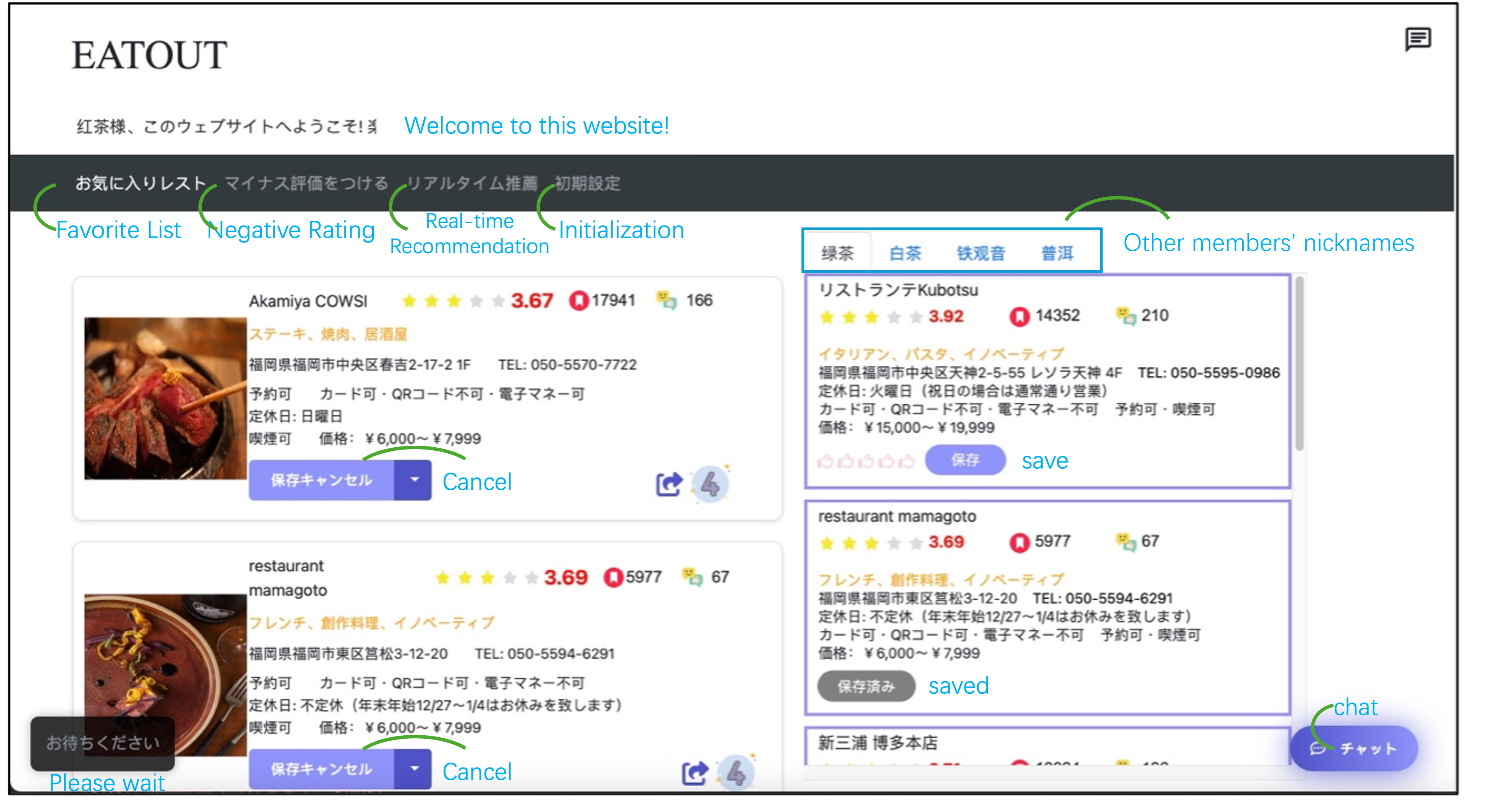}
    \caption{The ``Favorite List" Page.}
    \label{fig:shared_list}
\end{figure}

\item[c) Negative Rating:]
The {\it Negative Rating Page} allows users to rate restaurants they dislike (see Figure~\ref{fig:negative_rating}).
Only restaurants bookmarked by other members but not yet rated by the user are shown for evaluation.
The default score was $0$ if no rating was provided.

\begin{figure}[h!]
    \centering
    \includegraphics[width=0.9\textwidth]{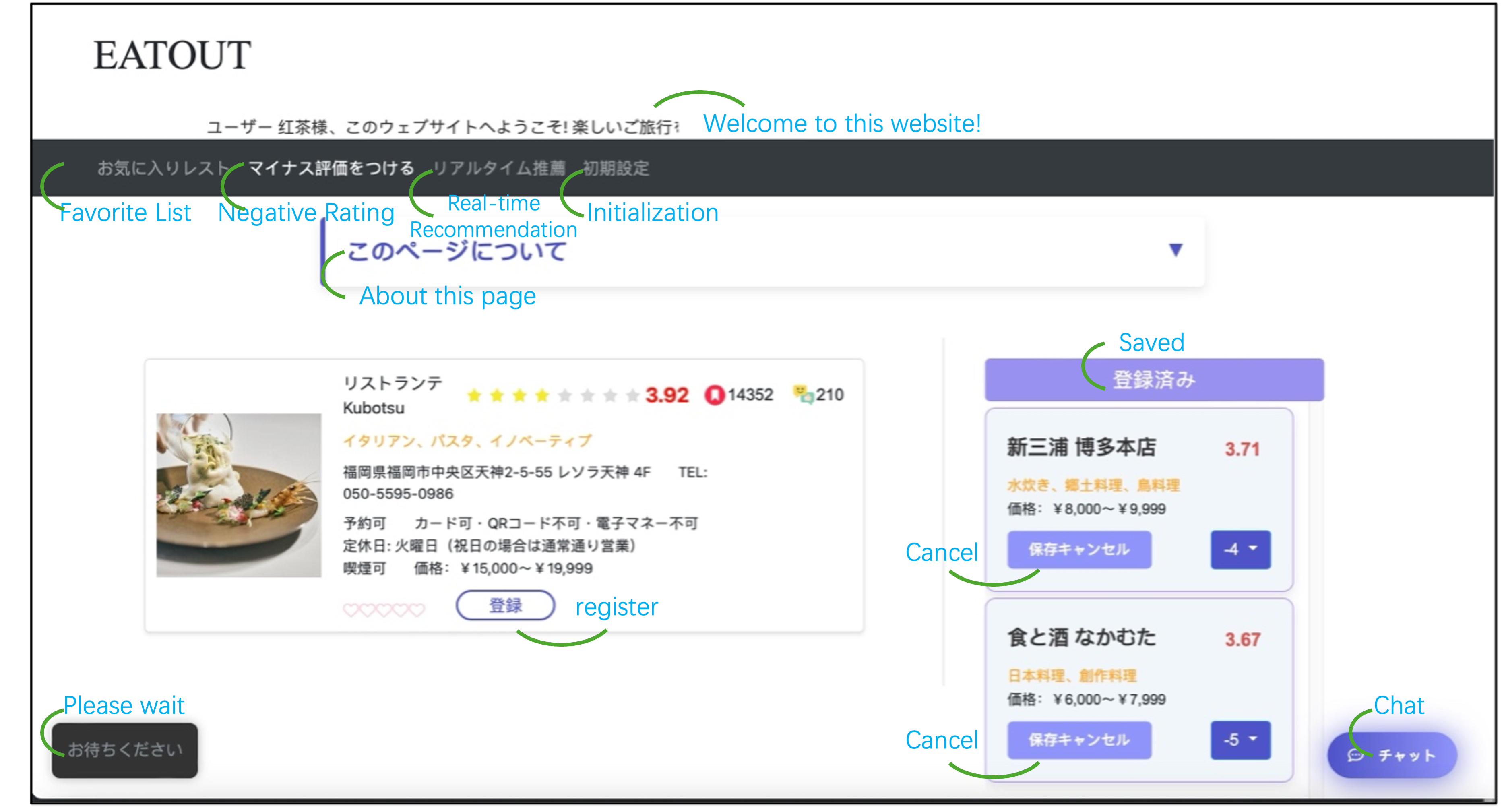}
    \caption{The ``Negative Rating" Page.}
    \label{fig:negative_rating}
\end{figure}

\item[d) Group Recommendation Results:]
The {\it Group Recommendation Page}  displays real-time recommendation results based on two algorithms (see Figure~\ref{fig:recommendation_results}):
\begin{description}
    \item[1) Baseline algorithm (IBGR):] Uses a trust calculation criterion focused on {\it leader influence} as described in the related work section.
    \item[2) Proposed algorithm:] Combines score similarity and trust degree by utilizing graph theory with the LeaderRank algorithm to generate recommendations.
\end{description}
Users can access the real-time recommendation results during discussions to facilitate ongoing group decision-making processes.

\end{description}

\section{Proposed Group Recommendation Algorithm}

We next discuss how we can leverage group interaction data to achieve 
effective group recommendation.  
%
%
%
%
%
We illustrate our proposed group recommendation algorithm in the context of
restaurant recommendation. 

In the {\it EATOUT} system\footnote{A custom-built system hosted on Amazon EC2.}, users provide initial restaurant ratings and bookmarks when they start using the system. The system then collect interaction data as users engage in group discussions. In particular, it records real-time chat logs, social bookmarking logs (including restaurant details, ratings, and sources), and negative ratings\footnote{They are recorded securely stored in a centralized MySQL database for subsequent analysis}. 
There are thus different types of data that can be 
exploited in making recommendations to groups:
\begin{description}
    \item[Initial Input Data:] Ratings and bookmarks provided by users during the initialization phase of the system. 
    \item[Interaction Data:] Real-time group interaction data including:
    \begin{description}
        \item[Chat Logs:] Messages exchanged during discussions by using the chat panel, which included both text-based messages and restaurant recommendations shared by participants in the form of links. 
        \item[Social Bookmarking Logs:] Actions where participants bookmarked restaurants from other users' preferred lists, including the restaurant bookmarked, the rating assigned to the restaurant, and the source of the bookmarked restaurant (i.e., whose list the restaurant was bookmarked from).
        \item[Negative Ratings:] Evaluations where participants gave minus scores to restaurants they found undesirable.
    \end{description}
\end{description}

\subsection{Overview of the Calculation Process}

\begin{figure}[h!]
    \centering
    \includegraphics[width=0.7\textwidth]{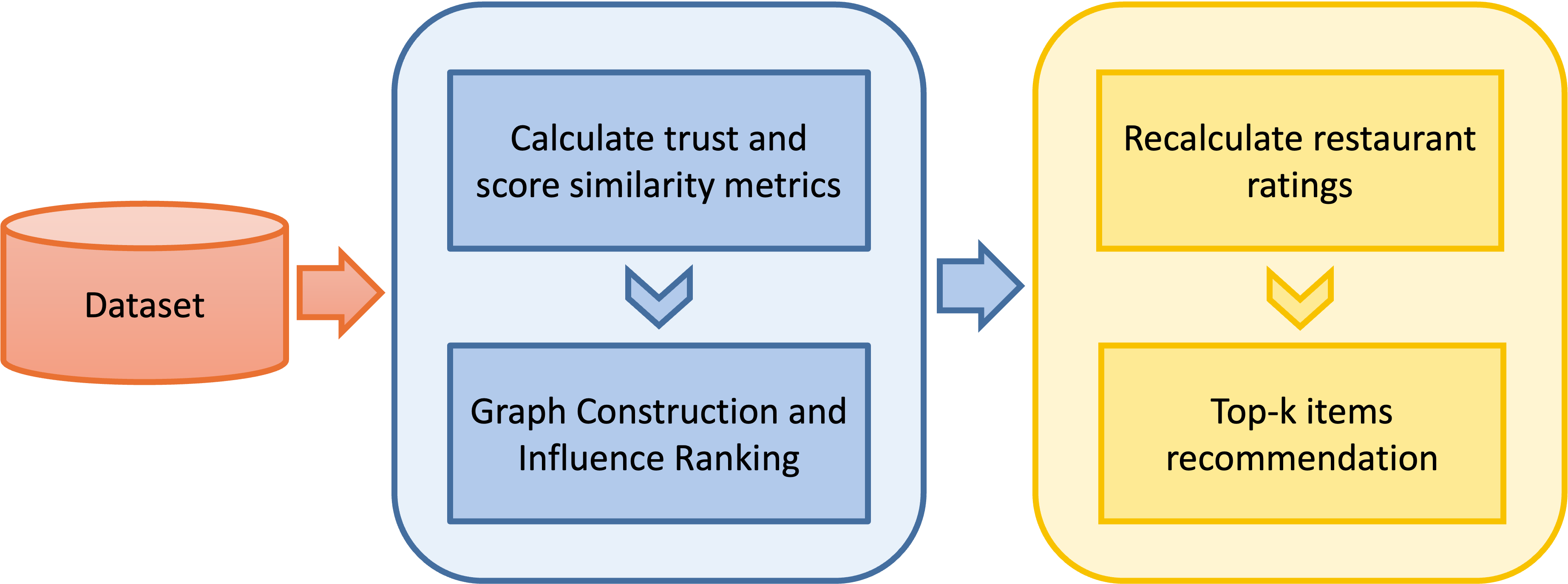}
    \caption{Block diagram of the proposed approach}
    \label{fig:block_diagram}
\end{figure}

The recommendation process is designed to combine individual preferences and group dynamics to produce recommendations that align with the collective interests of the group as shown in Figure~\ref{fig:block_diagram}. The process is structured into three main steps:

\begin{enumerate}
    \item \textbf{Rating Similarity and Trust Degree Calculation:}  
    The process begins by calculating the rating similarity and trust degree between each pair of group members. Rating similarity reflects the alignment of preferences between members based on their restaurant ratings, while trust degree quantifies the level of trust within the group, derived from interaction data and self-reported trust evaluations. These calculations result in two matrices that serve as inputs for subsequent steps.

    \item \textbf{Leader Identification Using LeaderRank:}  
    The LeaderRank algorithm~\cite{lu2011leaders} is applied to the trust and similarity matrices to determine the influence weight of each group member. This algorithm identifies the member with the highest influence weight as the group leader, who is considered the most impactful participant in the group decision-making process. LeaderRank incorporates both trust relationships and preference alignment to ensure a balanced representation of leadership.

    \item \textbf{Recalculation of Restaurant Ratings and Recommendation Generation:}  
    In the final step, the ratings of candidate restaurants are recalculated by incorporating the influence weights of all group members. The recalculated ratings are then ranked, and the top-3 restaurants with the highest scores are selected as the final recommendations for the group. This ensures that the recommendations reflect the collective preferences while giving appropriate consideration to the group leader's influence.
\end{enumerate}

\noindent
This calculation process integrates rating similarity, trust relationships, and leadership dynamics to generate recommendations that are both preference-aligned and socially contextualized.

\subsection{Rating Similarity and Trust Degree Calculation}
\subsubsection{Rating Similarity}
The rating similarity calculation is an essential step in identifying the alignment of preferences among group members. This process is based on the ratings provided by group members for a set of candidate restaurants.

\paragraph{Candidate Restaurants:}  
Each group member can select their preferred restaurants from the full list of available options, rate them on a scale from 1 to 5, and save them to their personal preferred list. Additionally, members have the option to view other group members’ preferred lists and assign negative ratings (ranging from -1 to -5) to restaurants they dislike. Restaurants without ratings are considered neutral. The union of all restaurants across the preferred lists of all group members is defined as the set of candidate restaurants.

\paragraph{Data Source:}  
The ratings for candidate restaurants are collected from all group members, including positive ratings, negative ratings, and unrated restaurants.

\paragraph{Pearson Correlation Coefficient (PCC):}  
The Pearson Correlation Coefficient (PCC) is used to calculate the similarity between the ratings of two group members. For each pair of members \(i\) and \(j\), the similarity is computed as:
\[
\text{Similarity}_{ij} = \frac{\sum_{k \in R} (r_{ik} - \bar{r}_i)(r_{jk} - \bar{r}_j)}{\sqrt{\sum_{k \in R} (r_{ik} - \bar{r}_i)^2} \sqrt{\sum_{k \in R} (r_{jk} - \bar{r}_j)^2}}
\]
where:
\begin{itemize}
    \item \( r_{ik} \): The rating given by member \(i\) to restaurant \(k\),
    \item \( \bar{r}_i \): The average rating of member \(i\),
    \item \( R \): The set of candidate restaurants rated by both members \(i\) and \(j\).
\end{itemize}

The resulting similarity scores are organized into a matrix, where each entry \(\text{Similarity}_{ij}\) represents the similarity between members \(i\) and \(j\). This matrix serves as the foundation for subsequent calculations, including trust degree evaluation and leader identification.

\subsubsection{Trust Degree}

Trust degree quantifies the level of trust among group members during group discussions, serving as a critical factor in leader prediction, which subsequently shapes the recommendation outcomes. It consists of two main components: {\it Chat-Based Trust} and {\it Save-Based Trust}.

\paragraph{Recipient Prediction Using SSA-GNN:}  
To compute chat-based trust, identifying the recipient of each message is essential. The Structure Self-Aware Graph Neural Network (SSA-GNN), proposed by Wang et al.~\cite{Wang2021ASS}, is employed to predict the recipient in multi-party dialogues. This model captures both local and global structural relationships, making it suitable for unstructured group discussions.

SSA-GNN constructs a fully connected discourse graph with their nodes  representing a sequence of EDUs (utterances) \( x_1, x_2, \dots, x_N \) from a dialogue, and edges representing structural relationships such as adjacency, speaker identity, and turn-taking patterns.

Edges \(r_{ij}^{(0)}\) between EDUs \((x_i, x_j)\) are initialized with learnable embeddings:
\[
r_{ij}^{(0)} = [s_{ij}, t_{ij}, d_{ij}]
\]
where:
\begin{itemize}
    \item \(s_{ij}\): Whether \(x_i\) and \(x_j\) share the same speaker,
    \item \(t_{ij}\): Adjacency indicator between \(x_i\) and \(x_j\),
    \item \(d_{ij}\): Relative position between \(x_i\) and \(x_j\).
\end{itemize}

Through iterative edge-centric message-passing, SSA-GNN refines edge and node embeddings, capturing discourse-level dependencies. The refined embeddings are then used to predict the recipient of each message by evaluating the likelihood of a specific recipient based on structural and semantic features.

Our approach employs
SSA-GNN to predict the recipient of each message during group discussions, providing critical input for the computation of trust degree. 
We integrated SSA-GNN into our system by treating contextual and language
issues as follows:

\begin{itemize}
    \item \textbf{Contextual Input:}  
    As SSA-GNN is a context-based model, we utilize a long polling mechanism to retrieve the latest conversations dynamically. For each new message, the previous \(n\) messages (where \(n = 5\)) are fetched as the context. This ensures that the model can consider both local and global dependencies within the dialogue.
    \item \textbf{Preprocessing:}  
    Since SSA-GNN accepts only English text as input, all messages are first translated to English from Japanese using a translation API. This step guarantees compatibility with the model while preserving the semantic meaning of the original messages.
\end{itemize}

\subsubsection{Chat-Based Trust}
Chat-based trust is derived from participants’ message exchanges during group discussions. It comprises two components: {\it Interaction Frequency} and {\it Sentiment Analysis}.

\paragraph{Interaction Frequency:}  
The interaction frequency quantifies the extent of interactions between two participants \(u\) and \(v\), incorporating a time-decay mechanism to give more weight to recent messages. It is defined as:
\[
\text{Trust}_{\text{chat}}^{\text{frequency}} = \frac{\sum_{i=1}^{N_{u \to v}} w_i}{\sum_{i=1}^{N_{u \to v}} w_i + \sum_{j=1}^{N_{v \to u}} w_j}
\]
where:
\begin{itemize}
    \item \(N_{u \to v}\): The total number of messages sent from \(u\) to \(v\),
    \item \(N_{v \to u}\): The total number of messages sent from \(v\) to \(u\),
    \item \(w_i\): The time-decay weight assigned to each message, defined as:
    \[
    w_i = e^{-\alpha \cdot (t_{\text{now}} - t_i)}
    \]
    \item \(t_{\text{now}}\): The current timestamp,
    \item \(t_i\): The timestamp of the \(i\)-th message,
    \item \(\alpha\): The decay factor, set to 0.01 in this study.
\end{itemize}

This formula ensures that recent messages contribute more to the interaction frequency, reflecting the dynamic nature of group discussions. The computed interaction frequency captures the proportion of \(u\)’s directed interactions towards \(v\) relative to their total interactions.

\paragraph{Sentiment Analysis:}  
Sentiment analysis evaluates the emotional tone of messages exchanged between participants. Using VADER (Valence Aware Dictionary and sEntiment Reasoner)~\cite{hutto2014vader}, the compound sentiment score \(S_{\text{sentiment}}(i)\) is computed for each message, representing its overall emotional intensity. A time-decay mechanism is applied to emphasize the influence of recent messages. The sentiment-based trust score is defined as:
\[
\text{Trust}_{\text{chat}}^{\text{sentiment}} = \frac{\sum_{i=1}^n w_i \cdot S_{\text{sentiment}}(i)}{\sum_{i=1}^n w_i}
\]
where:
\begin{itemize}
    \item \(n\): The total number of messages exchanged between \(u\) and \(v\),
    \item \(S_{\text{sentiment}}(i)\): The compound sentiment score of the \(i\)-th message,
    \item \(w_i\): The time-decay weight of the \(i\)-th message, defined as:
    \[
    w_i = e^{-\alpha \cdot (t_{\text{now}} - t_i)}
    \]
    \item \(t_{\text{now}}\): The current timestamp,
    \item \(t_i\): The timestamp when the \(i\)-th message was sent,
    \item \(\alpha\): The decay factor, set to 0.01 in this study.
\end{itemize}

The use of time-decay ensures that recent messages with strong emotional tones have a greater influence on trust calculations. Positive sentiment scores (\(S_{\text{sentiment}}(i) > 0\)) increase trust, while negative scores (\(S_{\text{sentiment}}(i) < 0\)) decrease it.

\paragraph{Time-Decay Mechanism:}  
Both interaction frequency and sentiment analysis incorporate a time-decay mechanism to reflect the temporal dynamics of group discussions. The weight \(w_i = e^{-\alpha \cdot (t_{\text{now}} - t_i)}\) ensures that recent interactions contribute more significantly to trust calculations, aligning with the evolving nature of group interactions. This approach captures not only the volume and emotional tone of interactions but also their temporal relevance.

\paragraph{Chat-Based Trust Aggregation:}  
The overall chat-based trust score is computed as the weighted combination of interaction frequency and sentiment-based trust scores. It is defined as:
\[
\text{Trust}_{\text{chat}} = \beta_1 \cdot \text{Trust}_{\text{chat}}^{\text{frequency}} + \beta_2 \cdot \text{Trust}_{\text{chat}}^{\text{sentiment}}
\]
where:
\begin{itemize}
    \item \(\text{Trust}_{\text{chat}}^{\text{frequency}}\): Trust score based on interaction frequency,
    \item \(\text{Trust}_{\text{chat}}^{\text{sentiment}}\): Trust score based on sentiment analysis,
    \item \(\beta_1\) and \(\beta_2\): Weighting parameters balancing the two components.
\end{itemize}

In this study, both weighting parameters are set to \(\beta_1 = \beta_2 = 0.5\), giving equal importance to interaction frequency and sentiment-based trust. This ensures that both the quantitative aspect of interactions (frequency) and their qualitative aspect (sentiment) are equally considered in the computation of chat-based trust.

\subsubsection{Saving-Based Trust}
The system provides a Favorite List feature, allowing users to save restaurants they are interested in. In addition to viewing their own favorite list, users can also access the favorite lists of other group members. This enables participants to explore restaurants selected by others and save the restaurants that they find appealing to their own list. Saving-based trust reflects the influence of participants’ preferences on each other, as inferred from the saving behavior. If a participant \(u\) saves a restaurant from another participant \(v\)’s  preferred list, it indicates \(u\)’s alignment with \(v\)’s preferences. Save-based trust is computed as:
\[
\text{Trust}_{\text{save}}(u, v) = \frac{\sum_{i=1}^k w_i \cdot S_{u, v}(i)}{\sum_{i=1}^k w_i}
\]
where:
\begin{itemize}
    \item \(k\): The total number of restaurants rated by both \(u\) and \(v\),
    \item \(w_i\): The weight reflecting the rating habit of participant \(u\), defined as:
    \[
    w_i = 1 + \frac{|u_i - \mu_u|}{\sigma_u}
    \]
    where:
    \begin{itemize}
        \item \(u_i\): The rating given by participant \(u\) to the \(i\)-th restaurant,
        \item \(\mu_u\): The mean rating of participant \(u\),
        \item \(\sigma_u\): The standard deviation of \(u\)’s ratings.
    \end{itemize}
    \item \(S_{u, v}(i)\): The rating similarity between \(u\) and \(v\) for the \(i\)-th restaurant, defined as:
    \[
    S_{u, v}(i) = 1 - \frac{|u_i - v_i|}{n_{\text{member}}}
    \]
    where:
    \begin{itemize}
        \item \(v_i\): The rating given by participant \(v\) to the \(i\)-th restaurant,
        \item \(n_{\text{member}}\): The total number of participants in the group.
    \end{itemize}
\end{itemize}

This formulation integrates two key aspects:
\begin{itemize}
    \item \textbf{Weight \(w_i\):} The weight reflects the relative importance of a rating based on the participant’s rating habit. Higher weights are assigned to ratings that deviate significantly from the participant’s mean rating (\(\mu_u\)), indicating that such ratings carry greater influence in determining trust. For example, when a participant gives a rating significantly higher than his/her usual rating habit, it implies a higher level of trust in the preferences of the other participant.
    \item \textbf{Similarity \(S_{u, v}(i)\):} The similarity score measures the degree of alignment between \(u\)’s and \(v\)’s ratings for a given restaurant. Scores closer to 1 indicate greater agreement in preferences.
\end{itemize}

\paragraph{Overall Trust Degree:}  
The overall trust degree between participants \(u\) and \(v\) is computed as the weighted sum of chat-based trust and save-based trust scores. It is defined as:
\[
\text{Trust}_{\text{degree}}(u, v) = \gamma_1 \cdot \text{Trust}_{\text{chat}}(u, v) + \gamma_2 \cdot \text{Trust}_{\text{save}}(u, v)
\]
where:
\begin{itemize}
    \item \(\text{Trust}_{\text{chat}}(u, v)\): Trust derived from chat-based interactions (including interaction frequency and sentiment analysis),
    \item \(\text{Trust}_{\text{save}}(u, v)\): Trust derived from save-based interactions (including rating similarity and user rating habits),
    \item \(\gamma_1\) and \(\gamma_2\): Weighting parameters balancing the contributions of chat-based and save-based trust.
\end{itemize}

In this study, both parameters are set to \(\gamma_1 = \gamma_2 = 0.5\), assigning equal importance to chat-based and save-based trust. This choice ensures that both interaction-based dynamics and preference alignment are equally reflected in the final trust degree.

\subsection{Leader Identification Using LeaderRank}

Leader identification is essential for determining the most influential participant within a group, as the leader’s preferences significantly impact the final recommendation results. In this study, we employ the LeaderRank algorithm to identify the leader, using a combined trust and rating similarity matrix to represent the relationships among group members.

\paragraph{Matrix Construction:}  
To apply the LeaderRank algorithm, a composite matrix \(M\) is constructed to capture both trust and rating similarity among group members. The matrix \(M\) is defined as:
\[
M = \lambda_1 \cdot M_{\text{rating\_similarity}} + \lambda_2 \cdot M_{\text{trust}}
\]
where:
\begin{itemize}
    \item \(M_{\text{rating\_similarity}}\): The rating similarity matrix, representing the alignment of ratings between participants,
    \item \(M_{\text{trust}}\): The trust matrix, representing the overall trust degree between participants,
    \item \(\lambda_1\) and \(\lambda_2\): Weighting parameters balancing the contributions of rating similarity and trust. In this study, both parameters are set to \(\lambda_1 = \lambda_2 = 0.5\), giving equal importance to both factors.
\end{itemize}

\paragraph{LeaderRank Algorithm:}  
The LeaderRank algorithm~\cite{lu2011leaders} is employed to calculate the influence scores of participants based on the composite matrix \(M\). The key steps are:

\begin{enumerate}
    \item \textbf{Graph Representation:}  
    construct a directed graph whose
    nodes represent group members 
    and edges, weighted by \(M_{vu}\), 
    capture relationships between members.

    \item \textbf{Score Initialization:}  
    Assign an initial score of \(1\) to all participants and a virtual ground node \(g\) :
    \[
    R^{(0)}(u) = 1, \quad R^{(0)}(g) = 1
    \]

    \item \textbf{Score Propagation:}  
    Update influence scores iteratively by
    redistributing scores along graph edges:
    \[
    R^{(t+1)}(u) = \sum_{v \in G} \frac{M_{vu} \cdot R^{(t)}(v)}{\sum_{w \in G} M_{vw}} + \epsilon \cdot R^{(t)}(g)
    \]
    where:
    \begin{itemize}
        \item \(M_{vu}\): Edge weight from participant \(v\) to \(u\),
        \item \(\epsilon\): Small constant ensuring graph connectivity.
    \end{itemize}

    \item \textbf{Final Redistribution:}  
    After convergence, redistribute the ground node’s score evenly among all participants:
    \[
    R(u) \gets R(u) + \frac{R(g)}{|G|}
    \]
\end{enumerate}

\paragraph{Leader Selection:}  
The participant with the highest influence score is identified as the leader:
\[
\text{Leader} = \arg\max_{u \in G} R(u)
\]
The identified leader is considered the most influential member of the group, as their preferences and interactions heavily shape the group’s decisions.

In this study, the composite matrix \(M\) integrates both rating similarity and trust to provide a comprehensive representation of the relationships among group members. By applying the LeaderRank algorithm, the system ensures that the leader is identified based on both preference alignment and interpersonal trust, thereby reflecting the dynamics of group decision-making more effectively.

\subsection{Recalculating Restaurant Ratings}

The recalculation process combines individual member ratings with their respective influence weights, derived from the LeaderRank algorithm, to compute group-level ratings for each candidate restaurant.

\paragraph{Recalculation Formula:}  
The group rating for each restaurant \(r_j\) is calculated as a weighted sum of individual ratings:
\[
\text{GroupRating}(r_j) = \sum_{u \in G} R(u) \cdot \text{Rating}_{u, j}
\]
where:
\begin{itemize}
    \item \(G\): The set of all group members,
    \item \(R(u)\): The influence score of group member \(u\),
    \item \(\text{Rating}_{u, j}\): The rating given by member \(u\) to restaurant \(r_j\).
\end{itemize}

\paragraph{Recommendation Generation:}  
Based on the recalculated group ratings, the system ranks all candidate restaurants. The top-\(k\) restaurants with the highest group ratings are selected as the final recommendations for the group:
\[
\text{Recommendation} = \text{Top-}k(\text{GroupRating})
\]
where \(k\) is the desired number of recommendations (e.g., \(k = 3\)).

\section{Experimental Design}
We conducted an experiment to evaluate the efficacy of our approach. This section outlines the participant demographics, experimental setup, and procedural steps we employed.

\subsection{Participants}
A total of 25 participants were recruited for the experiment, divided into 5 groups of 5 members each. Participants were university students aged between 20 and 29 years old, representing a variety of academic disciplines. Recruitment was conducted through campus announcements and online invitations. All participants were fluent in Japanese during the experiment and had prior experience with collaborative group activities. Individuals with conditions that could impair participation in discussions were excluded. 

As compensation for their time, each participant received a 1,000 yen gift card. Ethical approval for the study was obtained from the university's ethics committee, and all participants provided informed consent before the experiment commenced.

\subsection{Experimental Setup}

The experiment was conducted entirely online using a system deployed on an Amazon EC2 server, accessible via a public URL. The system was developed using Python's Flask framework for backend services and utilized a MySQL database for storing interaction data. The cloud-based infrastructure ensured a stable and uniform environment for all participants.

Before the experiment, participants attended an introductory session on Microsoft Teams, where they received detailed instructions on the objectives, procedures, and system functionalities. This session clarified any questions, ensuring consistent understanding of the tasks.

During the experiment, participants collaboratively selected the restaurant their group most preferred. The system recorded interaction data, including individual inputs, group discussions, decision-making sequences, and timestamps. These records were securely stored in the MySQL database for subsequent analysis.

After completing the tasks, participants filled out an online survey evaluating the system's usability, fairness of the group decision-making process, and overall satisfaction. This feedback provided insights into the effectiveness of the experimental design.

The use of Flask for backend development, MySQL for data storage, and standardized communication tools ensured a robust, controlled, and reproducible experimental environment.


\subsection{Procedure}
\begin{figure}[h!]
    \centering
    \includegraphics[width=0.7\textwidth]{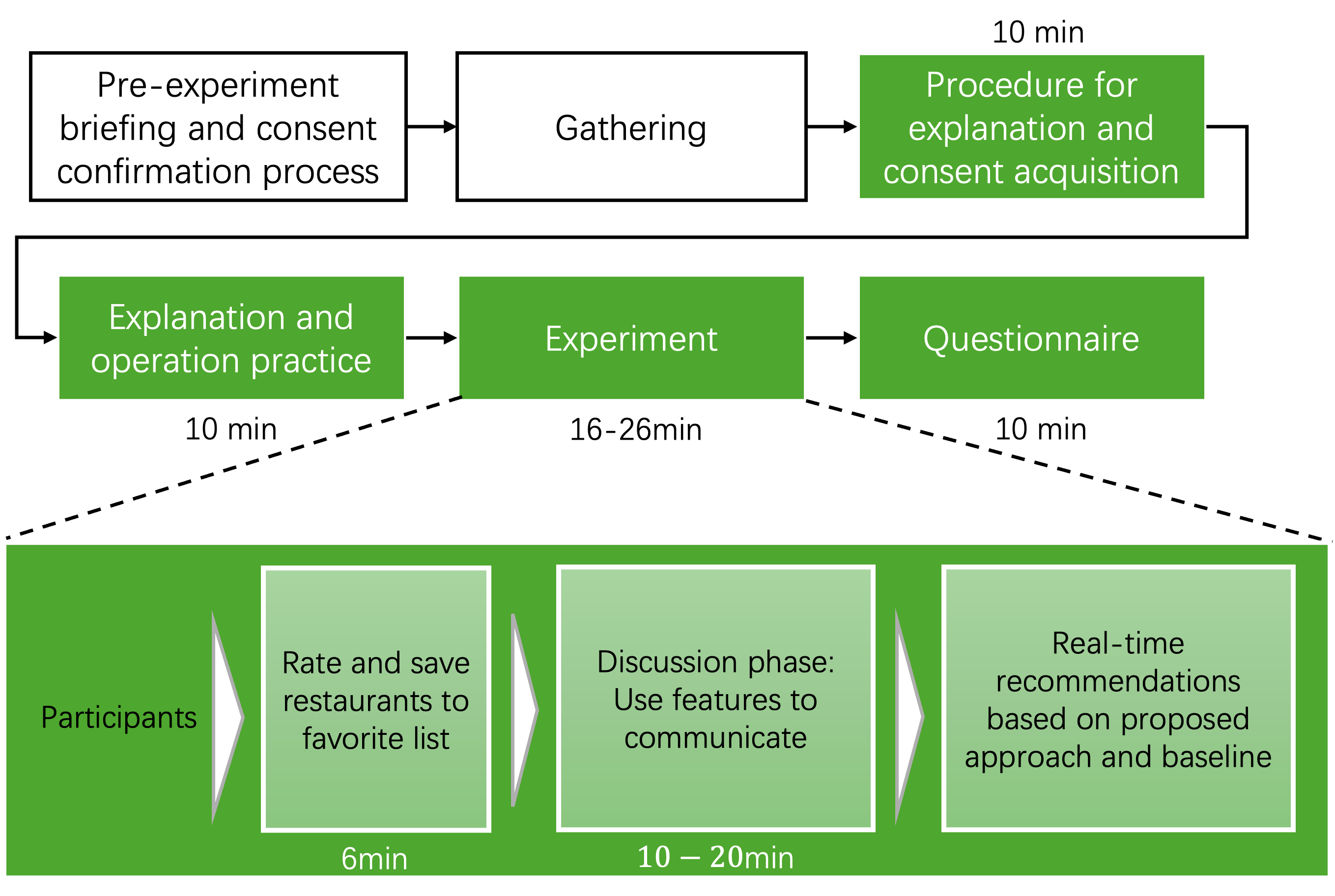} 
    \caption{Overview of the Experimental Procedure.}
    \label{fig:experiment_process}
\end{figure}

The experiment was conducted in a structured sequence of stages to ensure consistency and clarity. The process is described as follows.

To ensure privacy and confidentiality, all data were anonymized, and participants provided informed consent prior to the experiment. Collected data were securely stored on a centralized server for subsequent analysis.

\subsubsection{Pre-Experiment Briefing and Informed Consent}
At the start of the experiment, all participants were gathered in a virtual meeting conducted via Microsoft Teams. During this session, participants were provided with a comprehensive explanation of the experimental objectives, procedures, and their roles. This briefing included details about the tasks they would perform, the system's functionalities, and the data collection process. Participants were required to review and provide informed consent before proceeding to the subsequent stages.

\subsubsection{Introduction and Training}
Following the briefing, participants were introduced to the experimental system hosted on an Amazon EC2 server. A brief training session was conducted to ensure familiarity with the system's interface and interaction features. Participants were guided through key functionalities, including how to input ratings and utilize the interaction tools. This training ensured that all participants were adequately prepared for the tasks.

\subsection{Main Experiment}
The main experiment consisted of the following stages:

\subsubsection*{a) Welcome and Login}
Participants were welcomed with a page introducing the experiment's background, objectives, and process. After reading the instructions, they logged into the system using pre-assigned credentials provided during recruitment.

\subsubsection*{b) Restaurant Bookmarking and Initial Rating}
Participants bookmarked their favorite restaurants and rated various options within a 6-minute time limit. This step helped capture their initial preferences before group discussions.

\subsubsection*{c) Group Discussion and Decision-Making}
Participants engaged in a group discussion to select a restaurant collectively. The discussion lasted 10 to 20 minutes, and participants could use real-time chat and shared preference features to facilitate the process. 

\subsubsection*{d) Recommendation Assistance}
To aid decision-making, participants were provided with real-time recommendations based on two algorithms:
\begin{itemize}
    \item \textbf{Proposed Method}: Combines score similarity and trust degree, leveraging graph theory with the LeaderRank algorithm.
    \item \textbf{Baseline Method (IBGR)}: Relies on leader influence for trust calculation.
\end{itemize}

\paragraph{Discussion Termination Criteria}
The discussion phase lasts for 10 to 20 minutes. The system is equipped with a feature that automatically determines whether the discussion phase can be ended. This feature activates 10 minutes after the discussion begins. If 20 minutes elapse, the discussion phase is forcibly terminated.

\subsection{Post-Experiment Questionnaire}
After completing the discussion phase, participants answered a questionnaire that included subjective evaluations of the group dynamics, system usability, and satisfaction with the decision-making processes. It also included trust evaluations, where participants rated their trust levels for other group members.

\section{Result}

This section shows the result of the experiment, which will be separated into 4 sections. The result consists of data from both the questionnaire and the interaction data recorded by the system during the discussion.

\subsection{Leadership and Trust}
Since the proposed approach places an emphasis on the influence of the leadership, to evaluate the effectiveness of our approach on the aspect of the prediction of the leader and trust, as well as the standards of leader identification. 
For the prediction of the group leader, the question asks participants to choose one of the group members that they think plays the role of the leader during the discussion. The member who gets the highest approval is regarded as the questionnaire-based leader. The result is shown in the Table \ref{tab:leader_scores_compact}. It can be observed that in the first three groups, the leaders identified by the proposed approach align with the results of the questionnaire. In contrast, the baseline achieves consistency with the questionnaire results in only two groups. Notably, in both Group 1 and Group 3, all three methods produce consistent leader results, demonstrating a high level of agreement across approaches.

\begin{table}[h!]
\centering
\renewcommand{\arraystretch}{1.1} 
\setlength{\tabcolsep}{3pt}       
\small 
\begin{tabular}{|c|c|c|c|c|c|c|c|c|}
\hline
\textbf{Group} & \multicolumn{5}{c|}{\textbf{Leader Score}} & \textbf{Predicted} & \textbf{Baseline} & \textbf{Questionnaire} \\ 
               & \textbf{u1} & \textbf{u2} & \textbf{u3} & \textbf{u4} & \textbf{u5} & \textbf{Leader} & \textbf{Leader} & \textbf{Leader} \\ \hline
1              & 0.0 & 0.2 & 0.6 & 0.2 & 0.0 & u3 & u3 & u3 \\ \hline
2              & 0.8 & 0.0 & 0.2 & 0.0 & 0.0 & u1 & u4 & u1 \\ \hline
3              & 0.2 & 0.8 & 0.0 & 0.0 & 0.0 & u2 & u2 & u2 \\ \hline
4              & 0.0 & 0.8 & 0.2 & 0.0 & 0.0 & u3 & u4 & u2 \\ \hline
5              & 0.0 & 0.2 & 0.4 & 0.4 & 0.0 & u2 & u1 & u3, u4 \\ \hline
\end{tabular}
\caption{Leader Scores and Leadership Predictions Across Groups.}
\label{tab:leader_scores_compact}
\end{table}

Since we regard trust degrees as one of the factors for leader identification, the question asks participants 
to rate the trust degree towards other group members on the scale of 1-5 and 
compare leader and the person who gets the highest trust score in each group. 
To find out the person who gets the highest trust score, we take the average of the sum of the trust scores of other members for this member as the trust degree of this member. As shown in Table 5.2, We will find that in the five groups of data, the leader voted by group members and the most trusted person in each group are often the same. Based on the data from the questionnaire and the proposed approach, we can find that the consistency rates are all 80\%. 

\begin{table}[h!]
\centering
\begin{tabular}{cccc}
\toprule
\textbf{Group} & \textbf{Leader} & \textbf{Top User(Questionnaire)} & \textbf{Top User(Proposed)} \\
\midrule
1 & u3       & u3       & u3       \\
2 & u1       & u1       & u1       \\
3 & u2       & u3       & u2       \\
4 & u2       & u2       & u5       \\
5 & u3, u4    & u4, u5    & u3       \\
\bottomrule
\end{tabular}
\caption{Comparison of Leader, Questionnaire Highest Score, and Proposed Highest Score by Group}
\label{tab:comparison}
\end{table}

\begin{figure}[htbp]
    \centering
    \includegraphics[width=\textwidth]{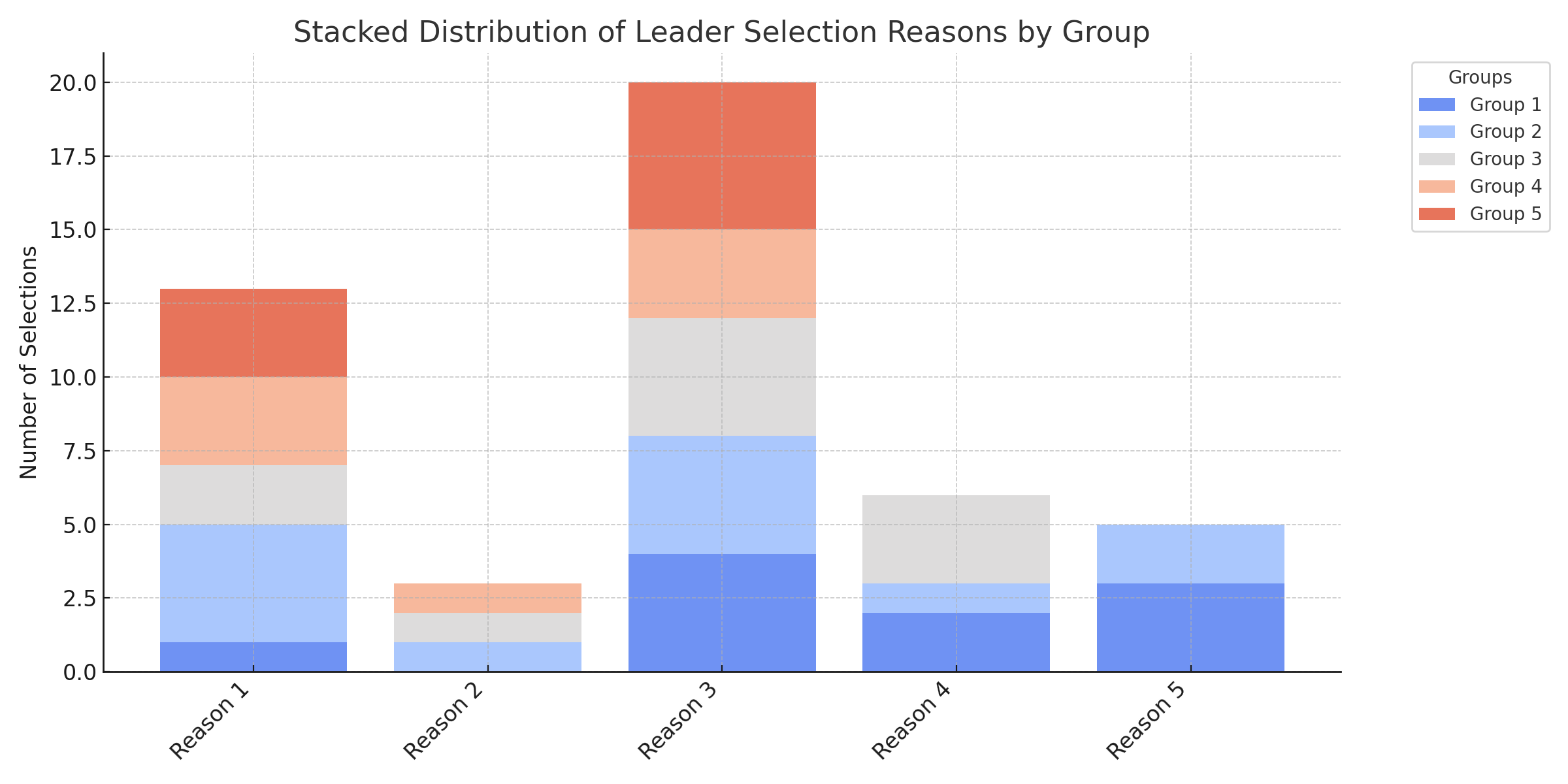}
    \caption{Stacked Distribution of Leader Selection Reasons by Group}
    \label{fig:leader_reasons}
\end{figure}

To find out the standards for leader identification, the question proposes several possible reasons for identifying a leader. Participants were asked to select all applicable reasons from the following options:
\begin{enumerate}
  \item {\it The person led the discussion and made most of the suggestions.}
  \item {\it Other members relied on or followed this person’s opinions.}
  \item {\it The person actively participated and responded to others' opinions.}
  \item {\it The person resolved disagreements and guided the group to consensus.}
 \item {\it Personally, the participant trusted this person’s judgment.}
\end{enumerate}
The result shown in the Figure \ref{fig:leader_reasons} shows that the first and the third reason get the highest votes because each group chose these two reasons.

\begin{figure}[htbp]
    \centering
    \includegraphics[width=\textwidth]{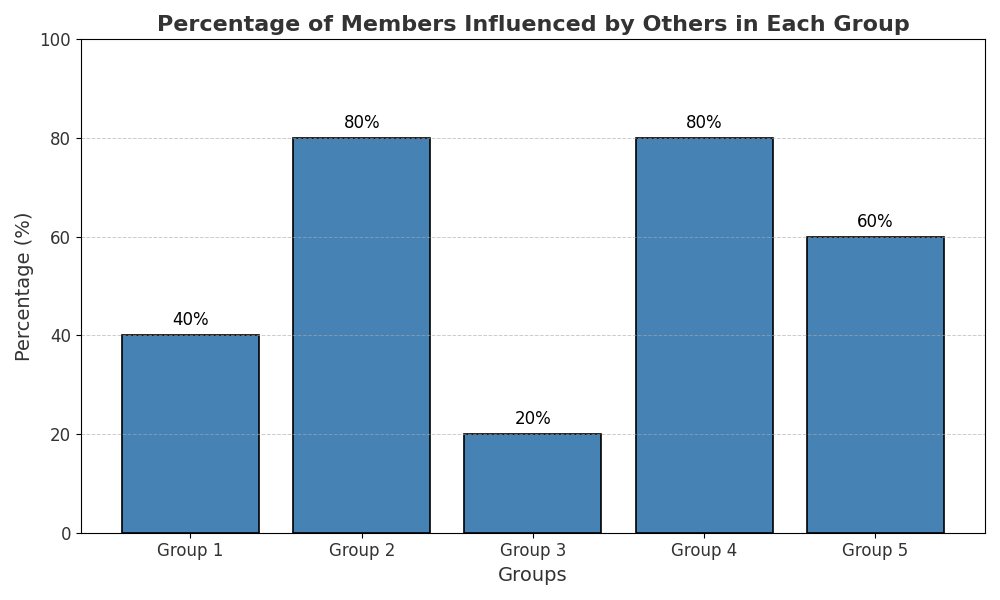}
    \caption{Percentage of Members Influenced by Others in Each Group}
    \label{fig:influnced_percentage}
\end{figure}

Finally, in this study, it is considered that a leader is the member who influences other members' opinions. For this hypothesis, the question asks participants whether they are influenced by others and who they think is the one who influences their opinions. 
The result is shown in Figure \ref{fig:influnced_percentage}. We can find that not all groups were very supportive of this hypothesis. Also only in Group 1 and Group 2, the person who influences other members' opinions is the same as the leader. In other 3 groups, although some people voted for the leader, the one who gets the hightest vote of influencing others‘ opinions is not the leader.

\subsection{Satisfaction with the Recommendation Results}

\begin{figure}[htbp]
    \centering
    \includegraphics[width=\textwidth]{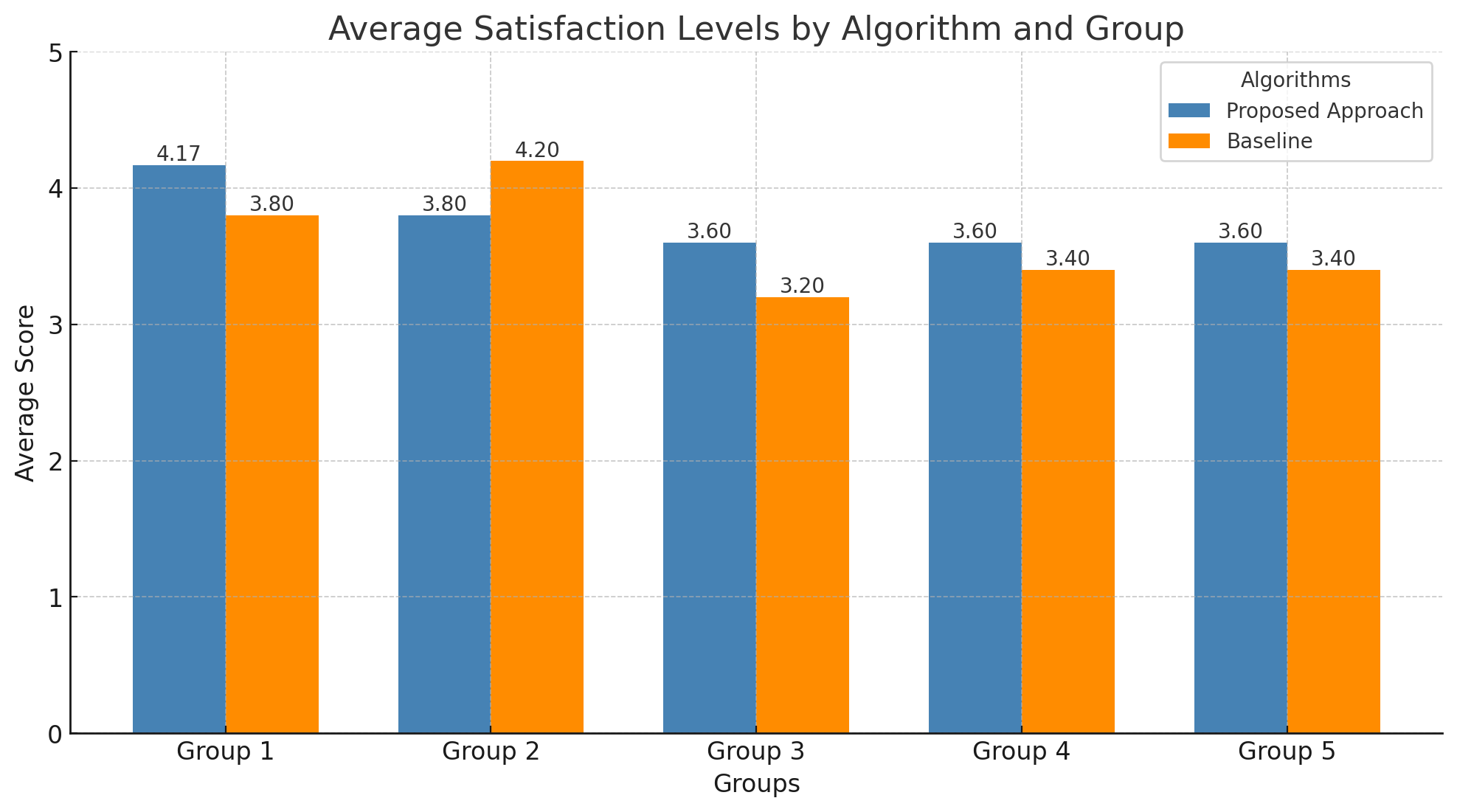}
    \caption{Average Satisfaction Levels}
    \label{fig:recommendation_satisfaction}
\end{figure}

To evaluate the performance of the proposed approach, the system provides 2 recommendations that each consists of 3 restaurants. The first one is based on the proposed approach, the other one is the baseline(IBGR), which I have introduced in the related work. The question asks participants to rate these 2 recommendations on the scale of 1-5. The result is shown in Figure~\ref{fig:recommendation_satisfaction}. We can find that in four out of five groups, the proposed approach has a better performance than the baseline algorithm. 
To find out the reason why the baseline algorithm has a better performance in Group 2, we analyzed the feedback on the recommendation results provided in the questionnaire of Group 2. Although we didn't find any reasonable answers, other groups' answers may provide some inspirations. The first aspect is that the dissatisfaction with a particular recommendation result stems from the appearance of restaurant the individual dislikes. The second aspect is that the inconsistency between the discussion content and the recommendation results. It also happened that members in one group have opposing opinions with the recommendation results. 


\begin{table}[ht]
\centering
\caption{Mean Satisfaction Scores and Standard Deviations for Algorithms Across Groups}
\label{tab:algorithm_satisfaction}
\begin{tabular}{@{}lccc@{}}
\toprule
\textbf{Group} & \textbf{Algorithm} & \textbf{Mean} & \textbf{Std. Dev.} \\ \midrule
1        & Proposed Approach        & 4.20          & 0.84               \\
               & Baseline        & 3.60          & 1.12               \\
2        &  Proposed Approach        & 3.60          & 0.89               \\
               & Baseline        & 3.80          & 0.75               \\
3        &  Proposed Approach        & 3.00          & 1.41               \\
               & Baseline        & 3.60          & 1.02               \\
4        &  Proposed Approach        & 3.80          & 0.75               \\
               & Baseline        & 3.40          & 0.89               \\
5        &  Proposed Approach        & 3.80          & 0.75               \\
               & Baseline        & 3.40          & 1.02               \\ \bottomrule
\end{tabular}
\end{table}

Table~\ref{tab:algorithm_satisfaction} presents the mean satisfaction scores and standard deviations for proposed approach and baseline across five participant groups. The standard deviations reflect the variability in user responses.

Proposed approach consistently received higher mean scores, particularly in Group 1 (4.20). However, its performance in Group 3 was less favorable, with the lowest mean score (3.00) and the largest variability (1.41), indicating significant user disagreement.

Baseline performed best in Group 2 (mean: 3.80; std. dev.: 0.75), demonstrating both high satisfaction and consistent evaluations. Conversely, its performance in Group 5 was the weakest (mean: 3.40), accompanied by notable variability (std. dev.: 1.02). Group 1 showed the most consistent satisfaction for Algorithm 1, while Group 3 exhibited the greatest variability for both algorithms.

\subsection{Satisfaction with the User Interface}
To investigate the satisfaction with the user interface of participants, the questionnaire included the following questions. Participants were asked to rate their agreement with the following statements on a scale of 1 to 5 (1 = Strongly disagree, 5 = Strongly agree).

\begin{table}[htbp]
\centering
\caption{Survey Questions}
\begin{tabular}{@{}cl@{}}
\toprule
\textbf{No.} & \textbf{Question} \\ \midrule
1 & The labels (buttons, menus, etc.) on the user interface are clear. \\
2 & The layout of the user interface is appropriate. \\
3 & The information provided about the restaurants is sufficient. \\
4 & I quickly got accustomed to using the recommendation system. \\ \bottomrule
\end{tabular}
\label{tab:survey_questions}
\end{table}

\begin{figure}[htbp]
    \centering
    \includegraphics[width=\textwidth]{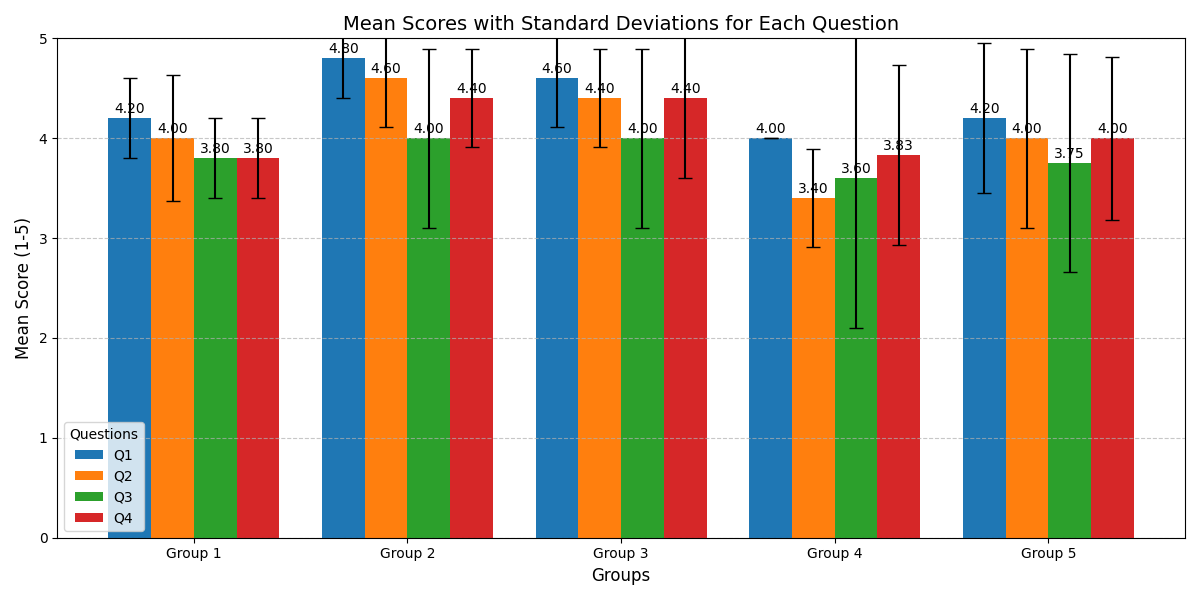}
    \caption{Satisfaction with UI}
    \label{fig:UI_satisfication}
\end{figure}

Figure~\ref{fig:UI_satisfication} illustrates the mean scores and standard deviations for each question across the five groups. The mean scores range from 3.4 to 4.8, indicating varied levels of user satisfaction with different aspects of the recommendation system.

Q1 (Interface Label Clarity) and Q4 (Ease of Use) consistently received high ratings with minimal variability, demonstrating the system's robustness in this area. Despite being a critical feature, Q3 consistently received lower mean scores across almost all groups (ranging from 3.6 to 3.83). Moreover, its standard deviation was the largest among all questions, indicating significant variability in user opinions. While some participants appreciated the information provided, others found it insufficient, likely due to unmet expectations regarding detail or diversity.

Furthermore, Q2 (Appropriate layout) received high ratings in most groups but performed poorly in Group 4, where the mean score fell to 3.4, highlighting specific layout-related issues.
Group 2 exhibited the highest overall satisfaction, particularly for Q1 and Q4. Groups 2, 3, and 5 exhibited similar scoring distributions, reflecting consistent user experiences and feedback.

\subsection{Discussion Process}

\begin{figure}[htbp]
    \centering
    \includegraphics[width=\textwidth]{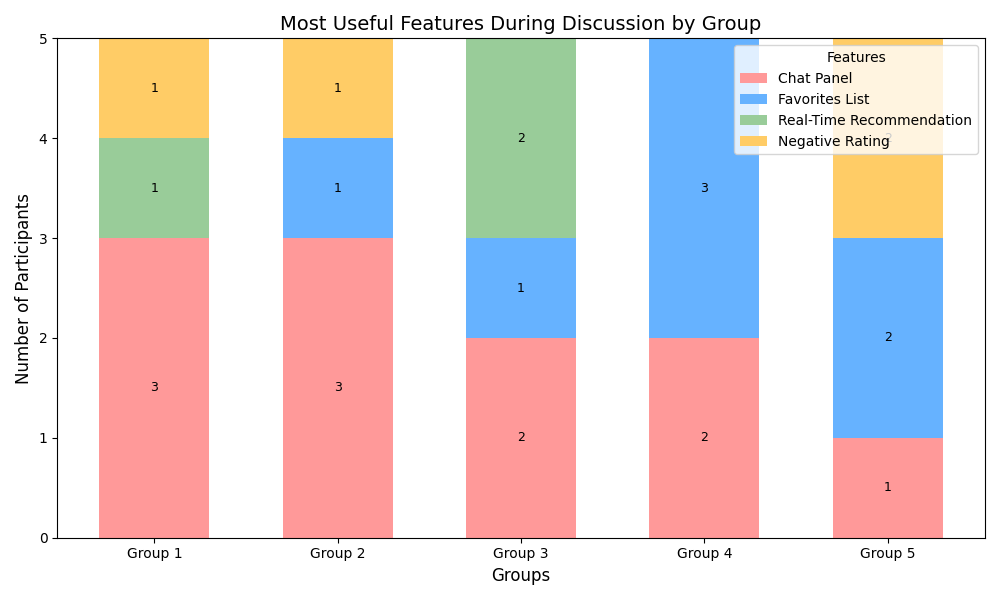}
    \caption{Votes for the Most Useful Functions}
    \label{fig:useful_features}
\end{figure}

As shown in the experiment setup, the system provides 4 functions during the discussion to facilitate communication. The data in Figure~\ref{fig:useful_features} reveals distinct preferences for discussion features across groups. The {\it Chat Panel} emerged as the most preferred feature overall, particularly in Group 1 and Group 2, where 3 participants each selected it. This indicates that real-time communication is crucial during discussions. Based on the reasons provided in the questionnaire, a commonly cited reason is that the chat panel enables the exchange of opinions, allowing users to understand others' preferences as well as the reasons behind their likes and dislikes. Additionally, the restaurant recommendation feature allows users to conveniently view and favorite restaurants recommended by other members.

The {\it Favorites List} was highly favored in Group 4 (3 participants) and Group 5 (2 participants), suggesting that these groups value referencing others' preferences. According to the questionnaire data, the reasons can be categorized into two aspects: the ability to refer to others' opinions and the opportunity to understand others' preferences. Particularly for the second reason, understanding others' preferences helps users selectively refer to those with similar interests.

In contrast, the {\it Real-Time Recommendation} feature had low overall selection rates, with Group 3 as an exception (2 participants), indicating a possible limited relevance or usability of this feature in its current form. The {\it Negative Rating} feature, while generally less selected, showed potential value in Group 5, where 2 participants chose it. Groups 1 and 2 showed minimal interest in this feature (1 participant each). According to the questionnaire data, a common reason is that users tend to have a clearer judgment about restaurants they dislike compared to those they like, thereby helping to avoid disliked restaurants in the final results.

Significant differences were observed among groups. Group 1 and Group 2 demonstrated balanced preferences, favoring the Chat Panel while showing little interest in other features. Group 4 strongly preferred the Favorites List, reflecting an emphasis on referencing other users' choices. Group 3 uniquely focused on Real-Time Recommendation, suggesting a higher demand for automated suggestions during discussions.

\begin{figure}[htbp]
    \centering
    \includegraphics[width=\textwidth]{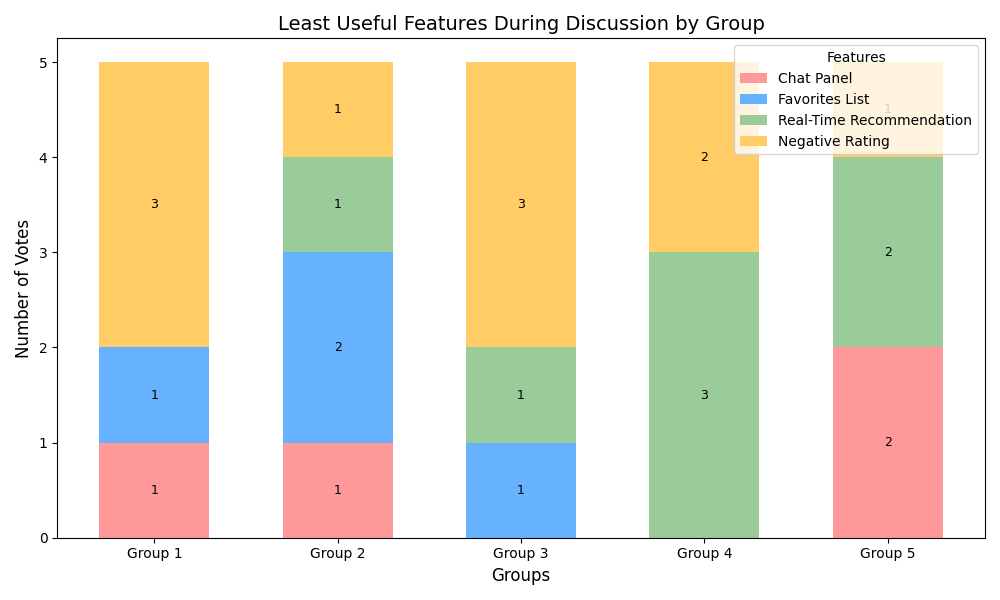}
    \caption{Votes for the Least Useful Functions}
    \label{fig:useless_features}
\end{figure}

On the other hand, Figure~\ref{fig:useless_features} shows the least useful functions during the discussion voted by each group. In Group 1 and Group 3, the {\it Negative Rating} feature was considered the least useful, receiving 3 votes in each group. Based on the reasons provided by users, the feedback can be categorized into the following aspects: First, users expressed that they did not receive meaningful feedback from their negative ratings on the recommendation results. Second, some users were concerned that their own or others’ negative ratings might adversely affect the final recommendation outcome. Third, users who had a high level of acceptance toward the recommended restaurants tended to disregard the negative rating feature altogether. Someone also thought that this function can be merged with the Favorites List.

Another feature that received high votes was the {\it Real-Time Recommendation}, particularly in Group 4, where it garnered three votes. The primary reason cited was the excessively high update frequency, with suggestions to reduce the frequency of updates. One participant mentioned that he placed greater trust in the conclusions reached through discussion compared to the recommendations provided by the system.

Although the {\it Chat Panel} received relatively low votes across all groups, some participants noted that relying solely on chatting is insufficient to influence opinions, especially among individuals who are not familiar with each other.

Additionally, some participants pointed out that the lack of sufficient restaurant information provided in the interface affected the effectiveness of some features.

   
\begin{figure}[htbp]
    \centering
    \includegraphics[width=\textwidth]{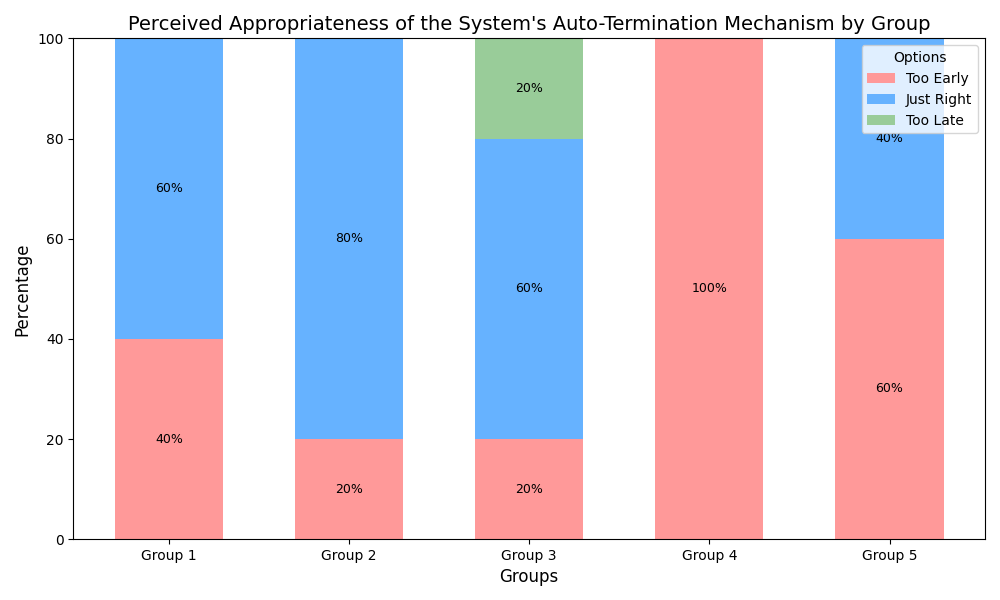}
    \caption{Perceived Appropriateness of the System's Auto-Termination Mechanism by Group}
    \label{fig:termination_appropriateness}
\end{figure}

Figure~\ref{fig:termination_appropriateness} shows the results of the questionnaire on the perceived appropriateness of the mechanism that terminates discussions automatically. It can be found that the first three groups generally perceived the discussion length as appropriate, while the latter two groups considered it to be too short. Especially in Group 4, all group members thought the termination mechanism is too early. The current mechanism is based on three criteria derived from the entropy of the score similarity matrix and the trust matrix. These criteria are evaluated every 30 seconds, starting 10 minutes into the discussion. If at least one criterion is satisfied for three consecutive evaluations, the system determines the discussion to be complete. However, during the actual discussions, instances were observed where the system prematurely concluded the discussion while participants were still actively engaged in the discussion. This indicates that the termination criteria may be overly lenient. 

The questionnaire results on the consensus achieved after discussion by group are shown in Figure~\ref{fig:termination_appropriateness}. It can be observed that the data for the last three groups is consistent, with 60\% of participants indicating partial consensus and 40\% reporting full consensus. In Group 2, all participants unanimously agreed that partial consensus was achieved. Notably, group 1 stands out, as 80\% of participants believed partial consensus was reached, but 20\% indicated that no consensus was achieved. 

\begin{figure}[htbp]
    \centering
    \includegraphics[width=\textwidth]{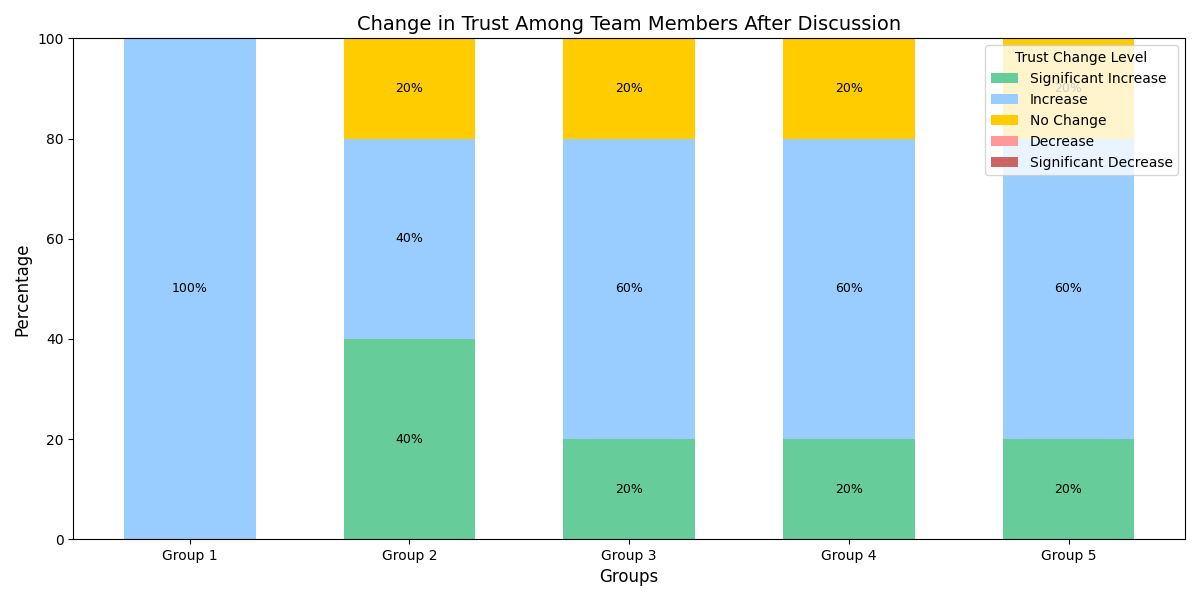}
    \caption{Change in Trust Among Team Members After Discussion}
    \label{fig:trust_change}
\end{figure}

The questionnaire results on the change in trust among team members after discussion are shown in Figure~\ref{fig:trust_change}. In Group 1, all members unanimously reported an increase in trust (100\%), with no participants selecting ``No Change" or ``Decrease.". While in group 2, opinions were more diverse in this group. While 40\% of participants reported a ``Significant Increase" and another 40\% indicated an ``Increase", 20\% of participants stated that there was ``No Change" in trust. The remaining groups demonstrated consistent distributions, with 60\% of members reporting an ``Increase," 20\% indicating a ``Significant Increase", and the remaining 20\% reporting ``No Change." This pattern reflects moderate but consistent improvements in trust among group members.

\begin{figure}[htbp]
    \centering
    \includegraphics[width=\textwidth]{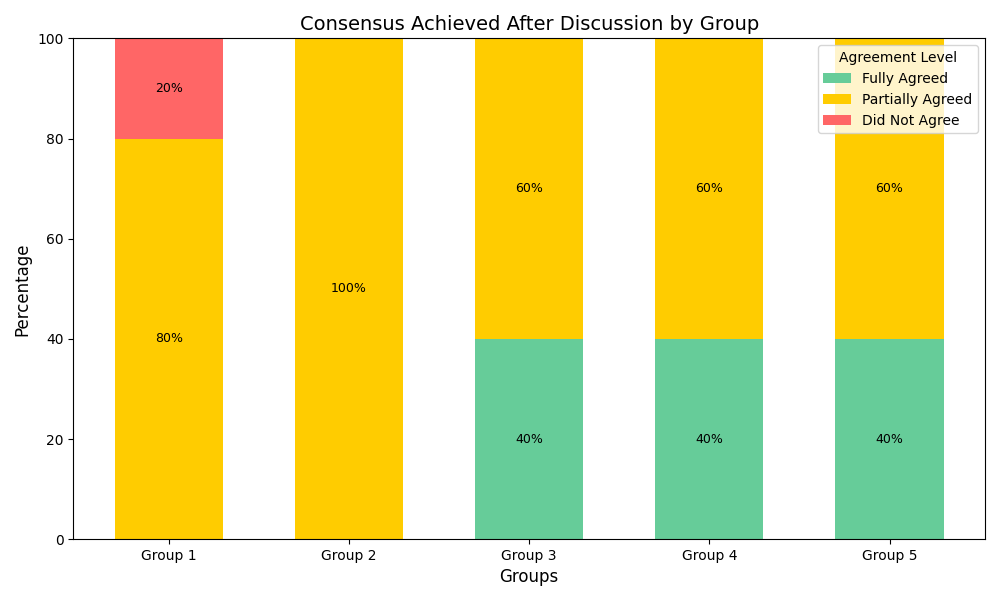}
    \caption{Consensus Achieved After Discussion by Group}
    \label{fig:consensus_achieved}
\end{figure}

\section{Discussion}
We next discuss the implications of the results and findings and explore the insights these results may offer for leader identification and user interface in group recommender systems. 

\subsection{Leader Identification and Influence}

In the algorithm proposed in this study, the pairwise relationships between members are considered based on trust levels and score similarity. Using graph theory, each member is treated as a node, and the pairwise relationship matrix is represented as directed edges, resulting in a directed graph. The LeaderRank algorithm is then applied to calculate the weight of each member and subsequently adjust their scores. 

The results indicate that the member with the highest weight assigned by the LeaderRank algorithm does not align well with the member receiving the highest number of votes in the questionnaire within each group. However, the member with the highest trust level, as derived from the trust matrix, shows a strong alignment with the member receiving the most votes in the questionnaire, achieving a consistency of 80\%. This confirms that the leader is often the most trusted member. Nonetheless, there may be room for improvement in the method of calculating trust degree and integrating trust levels and score similarity. 

When participants were asked about their criteria for identifying the leader, most of them preferred members who took a leading role in the discussion, provided suggestions, and actively participated by responding to others. These aspects primarily focus on the content of conversations during discussions. However, the calculation of trust levels in this study is based solely on the frequency of user interactions and sentiment intensity scores, without conducting a more detailed analysis of the textual content. Regarding speaking frequency, only pairwise relative frequencies were calculated, whereas users may perceive the overall number of contributions as more intuitive rather than the frequency of interactions with specific individuals. This mismatch could introduce potential biases. Furthermore, relying solely on the sentiment intensity of each utterance is insufficient to capture the context of discussion topics and their progression, which may also lead to inaccuracies.

Based on these observations, we can employ the following 2 approaches for enhancing the accuracy of leader identification ingroup recommmender systems:
\begin{enumerate}
    \item The proposed approach emphasizes the pairwise relationships between members. However, in team discussions, participants' perceptions are often shaped by overall contributions rather than detailed pairwise interactions. This divergence suggests that identifying a leader may also require consideration of each member’s overall performance within the group, alongside the pairwise relationships. Integrating both dimensions—pairwise relationships and individual contributions to the group as a whole—could enhance the accuracy of leader identification.
    \item The speaking frequency and sentiment scores collected by the system are not directly perceptible to participants. Additionally, individual ratings for restaurants tend to be overlooked as discussions progress. Ultimately, the discussion content itself plays a more significant role in team decision-making. Therefore, introducing large language models (LLMs) to analyze conversation content may offer an alternative approach. By evaluating each member's preferences toward specific discussion subjects (e.g., restaurants) and leveraging the real-time nature of conversations, it may be possible to replace individual restaurant ratings with discussion-driven evaluations. This approach could potentially improve recommendation satisfaction and decision-making effectiveness.
\end{enumerate}

The use of the LeaderRank algorithm in this study provided a calculated weight for each member within the team, which was subsequently used to recalculate the scores for each restaurant. Despite the algorithmic adjustments, questionnaire results revealed that only 20\% of participants in Group 3 believed that they were influenced by other members in their group. This indicates a potential gap between the algorithm's ability to model interpersonal influence and the actual perception of influence among participants. In Group 3, during the discussion, each member explicitly expressed their personal dietary restrictions or food preferences. The entire exchange was characterized by one member suggesting a type of food, followed by other members expressing their preferences or dislikes. It was evident that each participant had clear and distinct opinions. Members tended to agree if a suggestion aligned with their preferences and voiced their objections if it did not. This dynamics may explain why participants in this group perceived themselves as being less influenced by others. This suggests that in some group settings, decision-making can be more individualistic rather than collaborative, especially when participants have strong, well-defined preferences. The dynamics in Group 3 challenge algorithms like LeaderRank that rely on relational metrics, such as trust and interaction frequency, to model influence. When individual preferences dominate and interactions are limited to simple agreements or objections, the algorithm may fail to fully capture the nuanced group decision-making process.

\subsection{Proposed Approach vs. Baseline}
Regarding the comparison between the proposed approach and the baseline, the proposed approach incorporates a broader range of interaction data collected by the system. Unlike the baseline, which focuses solely on the number of shared preferences and score differences to calculate trust, the proposed approach also analyzes two additional interactive behaviors: chatting and bookmarking. This approach aims to uncover more latent trust relationships. Additionally, the proposed method employs the LeaderRank algorithm to integrate the trust matrix and the score similarity matrix, emphasizing pairwise relationships to calculate the weight of each member.

In contrast, the baseline identifies the leader by summing trust scores and similarity scores across all members, assigning leadership to the member with the highest cumulative score. It then calculates the leader's influence over other members to adjust the final restaurant scores. This method heavily relies on the influence of a single leader, assuming that decision-making follows a top-down structure where other members are primarily influenced by the leader.

The effectiveness of this leader-centric approach varies depending on the group’s decision-making structure. The baseline method is more suitable for groups where a single, dominant leader emerges, as it emphasizes the leader's influence over the entire group. Yet, in groups where leadership is distributed among multiple members or where no clear leader exists, this approach may struggle. In such cases, multiple members may have similar trust and similarity scores, but only the highest-scoring individual is designated as the leader, receiving a disproportionate influence in the recommendation process. This limitation makes the baseline less effective in scenarios where decision-making is decentralized, members actively contribute, and no single leader dominates the discussion.

Conversely, the proposed method is better suited for handling groups without a single leader or with multiple influential members. By emphasizing pairwise relationships rather than a singular leader’s impact, the proposed approach distributes influence more evenly, capturing the nuanced interactions among members. This makes it more adaptable to groups where multiple members lead discussions or where participants have strong individual opinions that are less influenced by others. By modeling trust and score relationships at a granular, member-to-member level, the proposed method ensures a more balanced and representative approach to let the group make consensus.

\subsection{User-Interface Design}
Based on the questionnaire results, one significant discussion point regarding feature design is the necessity of the negative rating function. As some participants noted, when making decisions, they often have a clearer idea of what they dislike rather than what they prefer. This aligns with the original intention behind the implementation of this feature, distinguishing it from the Favorites List by allowing participants to eliminate unwanted options without revealing their preferences to other members.
However, another perspective emerged during the study: some participants were reluctant to use this feature out of concern that their negative ratings might adversely affect the final outcome. This hesitation reflects a consideration for others’ opinions and suggests a potential drawback of the feature.

This divergence highlights an interesting phenomenon within group dynamics: some members exhibit more individualistic tendencies, focusing on their own preferences, while others lean toward collectivism, prioritizing the group’s overall harmony and decision-making process. Understanding this dichotomy is crucial for optimizing feature design to accommodate both individualistic and collectivist behaviors, ensuring the system supports diverse user perspectives.

Additionally, the existing negative rating feature has certain limitations, as its impact is not clearly reflected in the recommendation results. The current algorithm aggregates both positive and negative ratings, which may dilute the influence of negative feedback. A potential improvement would be to analyze negative ratings separately, identifying the categories of restaurants that members dislike. By doing so, these types of restaurants could be excluded from the recommendation results, thereby better aligning the system's output with user preferences.

\subsection{Limitation and Future Work}

In this study, while the proposed approach generally outperformed the baseline in terms of user satisfaction and demonstrated better adaptability in individualistic teams by emphasizing pairwise relationships over solely considering the leader's influence, there are still areas for improvement, including:

\begin{enumerate}
\item {\it Algorithm.}
The trust calculation could be enhanced by incorporating large language models (LLMs) to analyze conversation content. This approach could identify the discussed objects (e.g., restaurants or types of preferred/disliked food) and assess members’ sentiments toward them. By integrating pairwise relationships with each member's overall contribution to the group, the algorithm could better align with participants' natural thought processes. Additionally, analyzing members' negative ratings could help identify disliked food categories, thereby reducing the probability of such categories appearing in the final recommendations.

\item {\it Interface Design.} 
The interface could be improved by providing more detailed restaurant information, such as images and other visual elements, to help participants make more informed decisions and enhance user experience.

\item {\it Termination Mechanism.} 
Based on participant feedback, the discussion duration was considered too short in some groups, and most participants indicated that only partial consensus was reached after the discussion. The termination mechanism was primarily designed based on the changes in the entropy of trust and score similarity matrices, with the initial intent of ending discussions when they no longer significantly influenced the final outcomes. However, for the later groups, discussions often started later and exhibited lower interaction frequencies. Under such conditions, relying solely on entropy-based analysis can prematurely conclude the discussion. Therefore, incorporating the analysis of conversation content as an additional criterion for termination is crucial to ensure a more accurate assessment of discussion progress and impact.

\end{enumerate}




\section{Conclusion}

This study proposed a novel approach to group recommendation, which incorporates trust and score similarity matrices to enhance recommendation accuracy and group satisfaction. 
Unlike traditional approaches that solely focus on individual ratings or leader influence, the proposed approach emphasizes pairwise relationships and the collective dynamics within the group. This approach ensures that recommendations align more closely with the preferences of all members while maintaining fairness and inclusivity.

The results demonstrate that the proposed algorithm outperforms the baseline in terms of user satisfaction, achieving higher consistency between trust-based rankings and participants’ perceptions of leadership within the group. Specifically, the trust matrix showed an 80\% alignment with user-voted leaders, suggesting that trust is a critical factor in collaborative recommendation contexts. However, the integration of both trust and score similarity could benefit from further refinement to address cases where group dynamics are more complex.

Future research could leverage advanced natural language processing models to analyze conversation content, capturing users’ preferences more holistically. Furthermore, incorporating negative feedback (e.g., explicitly disliked restaurants) into the recommendation process could help avoid undesired options and improve satisfaction. Expanding the system to include richer restaurant metadata, such as images and detailed descriptions, could also enhance user engagement and decision-making.

This work contributes to the development of group recommendation systems by demonstrating the importance of modeling trust and interaction dynamics. By drawing inspiration from group decision-making research, the {\it EATOUT} system offers a novel approach to addressing the unique challenges of group recommendation, paving the way for applications in various domains such as social dining, travel planning, teamwork, and collaborative learning.

\begin{acks}
We would like to thank Miju Sugimoto, Kousuke Ejima, and Luyang Feng for their asssistance with the experiment,
and the members of the internal review board for their ethical review and approval of this study.
This research was funded by JSPS KAKENHI grant number JP20H00622.
\end{acks}

\bibliographystyle{ACM-Reference-Format}
\bibliography{sample-base}


\appendix

\section{Details of the Discussion Termination Mechanism}

To ensure efficient group discussions, a mechanism was designed to determine the optimal time to conclude the discussion phase. This mechanism evaluates the entropy of two matrices: \(\text{Matrix}_{\text{trust}}\) and \(\text{Matrix}_{\text{similarity}}\), which represent the trust and similarity relationships among group members, respectively. The entropy values provide a measure of stability within the group interactions.

\paragraph{Calculation of Entropy:}
At the start of the discussion, the entropy of \(\text{Matrix}_{\text{trust}}\) (\(\text{Entropy}_{\text{trust}}\)) and \(\text{Matrix}_{\text{similarity}}\) (\(\text{Entropy}_{\text{similarity}}\)) are calculated. These entropies are then recorded every 30 seconds.

\paragraph{Termination Criteria:}
From the 10th minute of the discussion onward, the mechanism checks for stabilization in group dynamics. The discussion is terminated if any of the following conditions are satisfied for three consecutive recordings:
\begin{enumerate}
    \item \(\text{Entropy}_{\text{trust}} < \text{Entropy}_{\text{similarity}}\),
    \item The change in \(\text{Entropy}_{\text{trust}}\) and \(\text{Entropy}_{\text{similarity}}\) compared to the previous recording is below a predefined threshold:
    \[
    |\text{Entropy}_{\text{trust}}^{(t)} - \text{Entropy}_{\text{trust}}^{(t-1)}| < \epsilon, \quad |\text{Entropy}_{\text{similarity}}^{(t)} - \text{Entropy}_{\text{similarity}}^{(t-1)}| < \epsilon
    \]
    \item The average change in entropy for both \(\text{trust}\) and \(\text{similarity}\) across recordings is below a threshold:
    \[
    \frac{\sum_{i=1}^{N} |\text{Entropy}_{\text{trust}}^{(t-i)} - \text{Entropy}_{\text{trust}}^{(t-i-1)}|}{N} < \epsilon, \quad \frac{\sum_{i=1}^{N} |\text{Entropy}_{\text{similarity}}^{(t-i)} - \text{Entropy}_{\text{similarity}}^{(t-i-1)}|}{N} < \epsilon
    \]
\end{enumerate}

\paragraph{Implementation:}
The system automatically monitors these criteria during the discussion phase. If any condition is met, the discussion is concluded, and participants are notified. This mechanism seeks to ensure that the discussion ends at a point of sufficient stability, improving the efficiency of the group decision-making process.

\end{document}